\documentclass[a4paper,11pt]{article}

\usepackage[utf8]{inputenc}
\usepackage{amsmath,amssymb}
\usepackage{graphicx}
\usepackage{geometry}
\usepackage{array}
\usepackage[table,xcdraw]{xcolor}
\usepackage{multicol}
\usepackage{tabularx}
\usepackage{subcaption}
\usepackage{jhep}
\usepackage{physics}
\usepackage{float}
\usepackage{tikz}
\usepackage{tikz-feynman}
\tikzfeynmanset{compat=1.1.0}
\usepackage{mathrsfs}
\usepackage{mathtools} 
\usepackage{tikz}
\usepackage{tikz-feynman}
\tikzfeynmanset{compat=1.1.0}

\title{\boldmath Asymmetric dark matter from leptogenesis in type-III seesaw framework with modular $S_4$ symmetry}

\author{Abhishek\textsuperscript{1,*}, V. Suryanarayana Mummidi\textsuperscript{1,$\dagger$}}
\affiliation{\textsuperscript{1}Department of Physics, National Institute of Technology,\\
Tiruchirappalli-620015, India}

\emailAdd{413120051@nitt.edu,venkata@nitt.edu}

\abstract{
We present a unified framework for neutrino masses, baryogenesis, and dark matter based on a modular $S_4$ symmetry combined with a type-III  seesaw mechanism. All Yukawa couplings, CP phases, and flavor textures originate from a single complex modulus~$\tau$, whose vacuum expectation value controls both visible and dark sector dynamics. The same modular parameter fixes the neutrino mass matrix, determines the CP asymmetries driving resonant leptogenesis, and correlates the resulting baryon and dark matter abundances. A detailed numerical analysis shows that the model reproduces all neutrino oscillation data within the $3\sigma$ NuFIT~5.2 (2024) ranges for normal ordering, predicting $\delta_{\rm CP} \simeq \pm (150^\circ-180^\circ)$, $\sum m_\nu\simeq(0.06-0.08)~\mathrm{eV}$,
and an effective Majorana mass
$m_{\beta\beta} \simeq (0.3 - 4)\times 10^{-3}~\mathrm{eV}$,
testable in next-generation neutrinoless double-beta decay experiments. The same modular Yukawas yield resonantly enhanced CP asymmetries
$|\epsilon_{L,\chi}| \sim 10^{-9}-10^{-6}$ at
$M_\Sigma \sim 10^{7}~\mathrm{GeV}$,
successfully generating the observed baryon asymmetry
$\eta_B\simeq6\times10^{-10}$ and dark relic density
$\Omega_\chi h^2\simeq0.12$ without additional free parameters.
The predicted correlation $\Omega_\chi/\Omega_B\simeq5.4$ fixes the dark matter mass to $m_\chi\simeq0.1-2~\mathrm{GeV}$, consistent with all current constraints. This framework therefore realizes a fully predictive baryon$-$dark matter co-genesis, where the geometry of the modular symmetry links the origin of flavor, CP violation, and the cosmic matter asymmetry.

}
\begin{document}
\maketitle
\newpage
\section{Introduction}

Neutrino physics continues to challenge the completeness of the Standard Model (SM) by revealing phenomena that cannot be accommodated within its original structure.  
The discovery of neutrino oscillations by Super--Kamiokande~\cite{SuperK:1998jle}, SNO~\cite{SNO:2001kpb}, and KamLAND~\cite{KamLAND:2008eeq} firmly established that neutrinos possess tiny but nonzero masses, requiring new degrees of freedom beyond the SM.  
At the same time, cosmological observations from Planck 2018~\cite{Planck:2018vyg} and DESI 2024~\cite{DESI:2024wgf} determine the baryon and dark matter relic densities with high precision,
\begin{equation}
\Omega_B h^2 = 0.0224 \pm 0.0001, \qquad 
\Omega_\chi h^2 = 0.120 \pm 0.001, \label{eq:relics}
\end{equation}
yielding the striking ratio $\Omega_\chi / \Omega_B \simeq 5.4$.  
The similarity of these two quantities, despite their different microphysical origins, strongly hints at a common underlying mechanism linking baryogenesis and dark matter genesis.  
Within the SM, however, CP violation from the quark sector is insufficient by many orders of magnitude to account for the observed baryon asymmetry $\eta_B = 6.1\times10^{-10}$~\cite{Planck:2018vyg}, and no viable dark matter candidate exists.  
These facts motivate extensions of the SM that can simultaneously explain neutrino masses, CP violation, the baryon asymmetry of the Universe (BAU), and the relic abundance of dark matter---ideally from a single origin.

\medskip

A particularly compelling approach to this problem is the asymmetric dark matter (ADM) hypothesis~\cite{Zurek:2013wia,Petraki:2013wwa}, in which the relic density of dark matter arises from a particle--antiparticle asymmetry analogous to the baryon asymmetry in the visible sector.  
If both asymmetries originate from a common process that violates lepton or baryon number in the early Universe, their comparable abundances can be naturally explained.  
This idea is realized in baryon--dark matter co--genesis models~\cite{Falkowski:2011zr,Patel:2022vjg,Biswas:2019ygr}, where visible and hidden asymmetries are generated simultaneously.  
Among the possible mechanisms, leptogenesis remains one of the most attractive, as it connects the cosmic matter asymmetry with the origin of neutrino masses~\cite{Davidson:2002qv,Buchmuller:2005eh}.

\medskip

In standard thermal leptogenesis~\cite{Fukugita:1986hr,Buchmuller:2005eh}, the CP--violating, out--of--equilibrium decays of heavy Majorana neutrinos create a lepton asymmetry that is partially converted to a baryon asymmetry through electroweak sphalerons.  
While this framework is elegant, it typically requires very heavy right--handed neutrinos ($M_N \gtrsim 10^{10}$~GeV)~\cite{Davidson:2002qv}.  
By contrast, in the resonant leptogenesis scenario~\cite{Pilaftsis:2003gt,Dev:2017wwc}, a small mass splitting between nearly degenerate states can resonantly enhance the CP asymmetry, allowing successful baryogenesis even for masses in the range $M_N \sim 10^7$--$10^9$~GeV.  
In such cases, both the magnitude of the asymmetry and its flavor dependence are determined by the underlying Yukawa structure and CP phases.

\medskip

The connection between leptogenesis and dark matter has been explored through several mechanisms in which the same heavy neutrino decays produce asymmetries in both the visible and hidden sectors~\cite{Falkowski:2011zr,Patel:2022vjg,Biswas:2019ygr}.  
Depending on the structure of the portal interactions, the resulting dark matter typically lies in the GeV range, yielding the correct ratio $\Omega_\chi / \Omega_B \sim 5$.  
Nevertheless, most existing constructions invoke multiple arbitrary couplings, additional mediator fields, or fine-tuned parameters.  
A symmetry--based explanation in which the flavor structure, CP phases, and portal couplings all emerge from a common theoretical origin is still missing.

\medskip

A natural step toward such a unified description arises from modular flavor symmetries~\cite{Feruglio:2017rjo,Kobayashi:2023wfa}, a new paradigm that replaces the traditional flavon mechanism with modular forms.  
In this approach, Yukawa couplings are modular functions of a single complex modulus $\tau$, which transforms under finite modular groups such as $A_4$, $S_4$, or $A_5$.  
The vacuum expectation value (vev) of $\tau$ determines the entire flavor structure, while its complex phase provides an intrinsic source of CP violation.  
Modular invariant models have been shown to reproduce realistic neutrino mixing patterns consistent with NuFIT global data~\cite{Kobayashi:2019,Penedo:2019,Wang:2020}.  
However, their implications for leptogenesis and asymmetric dark matter have remained largely unexplored.

\medskip

In this work, we develop a modular $S_4$--invariant framework that realizes resonant co--genesis of baryons and dark matter through a common source of CP violation determined by the modulus $\tau$.  
The modular symmetry fixes all Yukawa couplings, linking the flavor structure of leptons directly to the parameters relevant for early--Universe asymmetry generation.  
The two quasi--degenerate fermionic triplets $\Sigma_{1,2}$ responsible for neutrino mass generation also drive the co--genesis process, establishing a direct connection between the neutrino and dark sectors.

\medskip

Our construction employs the Type--III seesaw mechanism~\cite{Foot:1988qv,Ma:1998dn}, in which the heavy states are fermionic $SU(2)_L$ triplets with zero hypercharge.  
Unlike the singlet neutrinos of the Type--I seesaw, the triplet fields couple directly to the electroweak gauge bosons, providing richer phenomenology and potentially testable signatures at colliders.  
The effective light--neutrino mass matrix is generated through the relation
\begin{equation}
M_\nu \simeq - M_D^T M_\Sigma^{-1} M_D, \label{eq:typIII}
\end{equation}
where $M_D$ denotes the Dirac mass matrix and $M_\Sigma$ the mass of the heavy triplets.  
The presence of $SU(2)_L$ interactions allows the triplet states to remain in equilibrium until temperatures close to their mass scale, ensuring the correct washout dynamics for resonant leptogenesis while naturally realizing a low seesaw scale ($M_\Sigma \sim 10^7$~GeV).  
Moreover, the same modular structure that fixes $M_D$ also determines the couplings of $\Sigma_i$ to the dark sector, thereby linking neutrino masses, leptonic CP violation, and dark matter co--genesis.

\medskip

\noindent
We solve the coupled Boltzmann equations describing the simultaneous evolution of lepton and dark asymmetries generated by $\Sigma_1$ decays~\cite{Pilaftsis:2003gt,Dev:2017wwc},
\begin{itemize}
\item $\Sigma_1 \to L H$ (lepton asymmetry $Y_{\Delta L}$) and $\Sigma_1 \to \chi \phi$ (dark asymmetry $Y_{\Delta\chi}$),
\item including resonant CP violation, washout, and $\Sigma$--mediated transfer effects,
\item with sphaleron conversion $Y_B = (28/79) Y_{\Delta L}$,
\item yielding the DM relic ratio $\Omega_\chi/\Omega_B = (m_\chi/m_p) |Y_{\Delta\chi}/Y_{\Delta L}|$~\cite{Kaplan:2009ag}.
\end{itemize}
All asymmetries are controlled by Re$(\tau)$, Im$(\tau)$ from the single modular parameter.

\medskip

The remainder of this paper is organized as follows.  
Section~\ref{sec:Model} introduces the modular $S_4$ structure and the construction of the Yukawa couplings and the derivation of the neutrino mass matrices.  
Section~\ref{sec:leptogenesis} presents the Boltzmann framework describing co--genesis, while Section~\ref{sec:num} contains the numerical analysis and benchmark predictions.  
Finally, Section~\ref{sec:res} summarizes our conclusions and outlines possible phenomenological implications, including collider and gravitational--wave signatures.

\section{Model}
\label{sec:Model}

We consider a framework governed by the SM gauge symmetry 
$SU(2)_L \times U(1)_Y$, extended by a modular flavor symmetry $S_4$ of level~3~\cite{Feruglio:2017rjo} and a discrete stabilizing symmetry $Z_2^{\mathrm{DM}}$~\cite{Kaplan:2009ag}. 
The modular symmetry governs the flavor structure of leptons, while $Z_2^{\mathrm{DM}}$ ensures the absolute stability of the dark matter candidate. 
In this construction, neutrino masses arise via a type-III seesaw mechanism~\cite{Foot:1988qv,Ma:1998dn}, and the baryon asymmetry is generated through leptogenesis~\cite{Fukugita:1986hr}. 
A common fermion triplet mediator connects the visible and dark sectors, providing a unified origin for both the lepton asymmetry and the dark matter relic abundance~\cite{Falkowski:2011zr}.

The choice of the modular group $S_4$ is motivated by its ability to reproduce realistic lepton-mixing patterns through modular forms of low weight, avoiding the need for additional flavon fields~\cite{Kobayashi:2023wfa}. Its vev $\langle\tau\rangle$ spontaneously breaks the modular invariance, fixing the modular forms $Y_i(\tau)$ that determine the Yukawa structures of both visible and dark sectors. 
For concreteness, we note that the self-dual 
point $\tau = \omega = e^{2\pi i/3}$ is known 
to yield hierarchical and stable Yukawa 
textures in modular invariant 
constructions~\cite{Feruglio:2022koo}. This serves as a 
theoretical motivation for the viable 
parameter space found in our analysis. 
However, $\tau$ is not fixed by an explicit 
stabilization potential; instead it is 
determined phenomenologically by requiring 
consistency with low-energy neutrino 
observables, as discussed in Section~4.1.
The modular breaking scale $\Lambda_{\mathrm{mod}} \sim 10^{10}\,\mathrm{GeV}$ lies above the seesaw mass scale $M_\Sigma \sim 10^{7}\,\mathrm{GeV}$, thereby establishing a clear hierarchy between flavor generation and neutrino mass formation.

The field content and their transformation properties under 
$SU(2)_L \times U(1)_Y \times S_4 \times Z_2^{\mathrm{DM}}$ 
are summarized in Table~\ref{tab:fieldsA}. 
The model introduces a fermionic $SU(2)_L$ triplet $\Sigma$, which plays a central role in neutrino mass generation and in mediating the visible–dark sector connection. 
The scalar doublets $H_u$ and $H_d$ are the Higgs fields responsible for electroweak symmetry breaking, while an additional scalar $\phi$ and a fermion $\chi$ constitute the dark sector. 
The fields $\phi$ and $\chi$ are odd under $Z_2^{\mathrm{DM}}$, while all other fields remain even, preventing dark matter decay and ensuring the stability of $\chi$~\cite{Kaplan:2009ag}.

The discrete parity $Z_2^{\mathrm{DM}}$ is assumed to be exact and unbroken by higher-dimensional or gravitational operators. 
Operators such as $(L H_u)\chi/\Lambda$ are forbidden since $\chi$ and $\phi$ are odd under $Z_2^{\mathrm{DM}}$ whereas all other fields are even. 
Loop-induced corrections respect this discrete symmetry, guaranteeing the stability of the dark fermion $\chi$ even in the presence of Planck-suppressed effects.
\begin{table}[htbp]
\centering
\renewcommand{\arraystretch}{1.2}
\setlength{\tabcolsep}{7pt}
\caption{Transformation properties of matter and Higgs fields under 
$SU(2)_L \times U(1)_Y \times S_4 \times Z_2^{\mathrm{DM}}$. 
The modular weights $k_I$ are indicated in the last row.}
\label{tab:fieldsA}
\begin{tabular}{cccccccccc}
\hline
Field & $L$ & $l_{R_1}$ & $l_{R_2}$ & $l_{R_3}$ & $\Sigma$ & $H_u$ & $H_d$ & $\phi$ & $\chi$\\
\hline
$SU(2)_L$           & $2$ & $1$ & $1$ & $1$ & $3$ & $2$ & $2$ & $3$ & $2$ \\
$U(1)_Y$            & $-1/2$ & $-1$ & $-1$ & $-1$ & $0$ & $+1/2$ & $-1/2$ & $+1/2$ & $-1/2$ \\
$S_4$               & $3$ & $1'$ & $1$ & $1'$ & $2$ & $1$ & $1$ & $1$ & $2$ \\
$Z_2^{\mathrm{DM}}$ & $+$ & $+$ & $+$ & $+$ & $+$ & $+$ & $+$ & $-$ & $-$ \\
$k_I$               & $-2$ & $0$ & $-2$ & $-2$ & $-2$ & $0$ & $0$ & $0$ & $-2$ \\
\hline
\end{tabular}
\end{table}
\begin{table}[htbp]
\centering
\renewcommand{\arraystretch}{1.2}
\setlength{\tabcolsep}{7pt}
\caption{Transformation properties of modular Yukawa couplings under 
$SU(2)_L \times U(1)_Y \times S_4 \times Z_2^{\mathrm{DM}}$. 
The modular weight $k_I$ corresponds to the form weight.}
\label{tab:fieldsB}
\begin{tabular}{ccccccc}
\hline
 & $Y^{(2)}_{3}$ & $Y^{(2)}_{3'}$ & $Y^{(4)}_{1}$ & $Y^{(4)}_{3}$ & $Y^{(4)}_{3'}$ & $Y^{(4)}_{2}$\\
\hline
$S_4$               & $\mathbf{3}$ & $\mathbf{3'}$ & $\mathbf{1}$ & $\mathbf{3}$ & $\mathbf{3'}$ & $\mathbf{2}$ \\
$k_I$               & $2$ & $2$ & $4$ & $4$ & $4$ & $4$\\
\hline
\end{tabular}
\end{table}

\noindent
The modular forms appearing in the Yukawa interactions transform as multiplets of $S_4$ and are listed in Table~\ref{tab:fieldsB}. 
They carry modular weights corresponding to the order of the modular function, which determines the allowed couplings in the Lagrangian. The symmetry-breaking chain of the framework is expressed as
\begin{equation}
S_4^{\mathrm{modular}} \times SU(2)_L \times U(1)_Y
\xRightarrow[\langle\tau\rangle]{\text{spontaneous modular breaking}}
SU(2)_L \times U(1)_Y
\xRightarrow[\langle H_{u,d}\rangle = v]{\text{EWSB}}
U(1)_{\mathrm{em}},
\label{eq:symmbreaking}
\end{equation}
where $\langle\tau\rangle$ generates the modular forms that fix the Yukawa structures, while the Higgs vev $v \simeq 246~\mathrm{GeV}$ triggers electroweak symmetry breaking.

\noindent

Each operator in the Lagrangian satisfies the modular weight conservation condition 
$\sum_i k_i + k_Y = 0$, ensuring modular invariance~\cite{Feruglio:2017rjo}. Explicitly,
\begin{align*}
(L l_{R_1})_{\mathbf{3'}} H_d Y^{(2)}_{\mathbf{3'}}: &\quad (-2) + (0) + (0) + (2) = 0,\\
(L \Sigma)_{\mathbf{3}} H_u Y^{(4)}_{\mathbf{3}}: &\quad (-2) + (-2) + (0) + (4) = 0,\\
(\Sigma \chi \phi) Y^{(4)}_{1}: &\quad (-2) + (-2) + (0) + (4) = 0.
\end{align*}

The complete Lagrangian of the model is written as
\begin{equation}
\mathcal{L} = 
\mathcal{L}_{\ell} +
\mathcal{L}_{D} +
\mathcal{L}_{\Sigma} +
\mathcal{L}_{\mathrm{DM}},
\end{equation}
where the terms correspond to the charged-lepton, Dirac-neutrino, Majorana, and dark-matter portal sectors, respectively. 

The charged-lepton Yukawa couplings consistent with the modular $S_4$ symmetry take the form
\begin{equation}
\mathcal{L}_{\ell} =
\alpha (L l_{R_1})_{\mathbf{3'}} H_d Y^{(2)}_{\mathbf{3'}} 
+ \beta (L l_{R_2})_{\mathbf{3}} H_d Y^{(4)}_{\mathbf{3}} 
+ \gamma (L l_{R_3})_{\mathbf{3'}} H_d Y^{(4)}_{\mathbf{3'}} 
+ {\rm h.c.},
\end{equation}
which, after electroweak symmetry breaking, leads to the charged-lepton mass matrix
\begin{equation}
M_\ell = v_d
\begin{pmatrix}
\alpha\,Y_3 &
-2\beta\,Y_2 Y_3 &
2\gamma\,Y_1 Y_3 \\[6pt]
\alpha\,Y_5 &
\beta \!\left( \sqrt{3}\,Y_1 Y_4 + Y_2 Y_5 \right) &
\gamma \!\left( \sqrt{3}\,Y_2 Y_4 - Y_1 Y_5 \right) \\[6pt]
\alpha\,Y_4 &
\beta \!\left( \sqrt{3}\,Y_1 Y_5 + Y_2 Y_4 \right) &
\gamma \!\left( \sqrt{3}\,Y_2 Y_5 - Y_1 Y_4 \right)
\end{pmatrix}.
\end{equation}

The Dirac neutrino mass term arises from
\begin{equation}
\mathcal{L}_D =
\alpha_D (L \Sigma)_{\mathbf{3}} H_u Y^{(4)}_{\mathbf{3}} 
+ \beta_D (L \Sigma)_{\mathbf{3'}} H_u Y^{(4)}_{\mathbf{3'}} 
+ {\rm h.c.},
\end{equation}
leading to the Dirac mass matrix $M_D$:
\begin{equation}
M_D = v_u\,
\resizebox{0.92\textwidth}{!}{$
\begin{pmatrix}
-2\alpha_D\,Y_2 Y_3 &
-2\beta_D\,Y_1 Y_3 \\[8pt]
-\alpha_D\!\left( \tfrac{\sqrt{3}}{2}Y_1 Y_4 + \tfrac{1}{2}Y_2 Y_5 \right)
+ \beta_D\!\left( \tfrac{3}{2}Y_2 Y_5 - \tfrac{\sqrt{3}}{2}Y_1 Y_4 \right)
&
\alpha_D\!\left( \tfrac{3}{2}Y_1 Y_5 + \tfrac{\sqrt{3}}{2}Y_2 Y_4 \right)
+ \beta_D\!\left( -\tfrac{3}{2}Y_1 Y_5 + \tfrac{\sqrt{3}}{2}Y_2 Y_4 \right)
\\[8pt]
-\alpha_D\!\left( \tfrac{\sqrt{3}}{2}Y_1 Y_5 + \tfrac{1}{2}Y_2 Y_4 \right)
+ \beta_D\!\left( \tfrac{3}{2}Y_2 Y_4 - \tfrac{\sqrt{3}}{2}Y_1 Y_5 \right)
&
\alpha_D\!\left( \tfrac{3}{2}Y_1 Y_4 + \tfrac{\sqrt{3}}{2}Y_2 Y_5 \right)
+ \beta_D\!\left( -\tfrac{3}{2}Y_1 Y_4 + \tfrac{\sqrt{3}}{2}Y_2 Y_5 \right)
\end{pmatrix}
$}.
\end{equation}

The heavy Majorana mass term for $\Sigma$ is given by
\begin{equation}
\mathcal{L}_\Sigma =
\frac{1}{2} M Y^{(4)}_1 \, {\rm Tr}(\Sigma^c \Sigma)
+ \frac{1}{2} M_\epsilon \left(\Sigma^c \cdot Y^{(4)}_2 \otimes \Sigma \right)_{\mathbf{1}}
+ {\rm h.c.},
\end{equation}
yielding
\begin{equation}
M_\Sigma =
\begin{pmatrix}
M\!\left(Y_1^2 + Y_2^2\right) - M_\epsilon\!\left(Y_2^2 - Y_1^2\right) &
M_\epsilon\,Y_1 Y_2 \\[6pt]
M_\epsilon\,Y_1 Y_2 &
M\!\left(Y_1^2 + Y_2^2\right) + M_\epsilon\!\left(Y_2^2 - Y_1^2\right)
\end{pmatrix}.
\label{eq:2.8}
\end{equation}
The eigenvalues of $M_\Sigma$ correspond to two quasi-degenerate heavy Majorana mass eigenstates $\Sigma_{1,2}$ with masses $M_{1,2}$, whose small modular-induced splitting enables resonant enhancement. We note that $M_\epsilon$ sets the overall 
mass scale of the heavy triplets, while $M$ 
controls the small mass splitting 
$\Delta M = 2M$ between the two eigenstates, 
naturally realizing the quasi-degenerate 
spectrum required for resonant leptogenesis 
despite the hierarchy $M_\epsilon \gg M$.

The light neutrino mass matrix is then obtained through the type-III seesaw relation,
\begin{equation}
M_\nu = - M_D M_\Sigma^{-1} M_D^{T},
\end{equation}
linking low-energy neutrino parameters directly to the modular structure. 

Finally, the dark sector interacts through the portal coupling
\begin{equation}
\mathcal{L}_{\mathrm{DM}} =
 y_{DM}\, \mathrm{Tr}\!\left[\, \Sigma (\chi \phi) \,\right] Y_2^{(4)} + {\rm h.c.},
 \label{eq:2.9}
\end{equation}
where $\chi$ is stabilized by the discrete symmetry $\mathbb{Z}_2^\text{DM}$,
and $y_\text{DM}$ denotes the dark-sector Yukawa coupling that controls
the strength of the portal interaction between the heavy triplet $\Sigma$
and the dark fields $(\chi, \phi)$. 
After the breaking of the modular $S_4$ symmetry, the modular form $Y_2^{(4)}(\tau)$ induces
a specific flavor structure for the dark Yukawa coupling, which can be written as
\begin{equation}
Y_\chi = y_{DM}
\begin{pmatrix}
Y_1^2 - Y_2^2 & 2 Y_1 Y_2 \\[4pt]
2 Y_1 Y_2 & -Y_1^2 + Y_2^2
\end{pmatrix}.
\label{eq:Ychi_matrix}
\end{equation}
This interaction transfers the generated lepton asymmetry to the dark sector
through $\Sigma$ exchange~\cite{Falkowski:2011zr},
resulting in correlated baryon and dark matter abundances.
All coefficients 
$\alpha,\beta,\gamma,\alpha_D,\beta_D,M,M_\epsilon \sim \mathcal{O}(1)$ 
are real free parameters that can be traded for rescaled modular forms 
$Y_i \to \alpha_i Y_i(\tau=\omega)$, whose values are fixed by requiring
realistic mass hierarchies consistent with oscillation data.

\section{Leptogenesis and Asymmetric Dark Matter}
\label{sec:leptogenesis}
In this section we describe a co-genesis scenario in which the BAU and the DM relic abundance originate simultaneously from the CP-violating decays of nearly degenerate heavy Majorana fermions~\cite{Falkowski:2011zr,Dutta:2011jh}. 
The framework relies on resonant leptogenesis~\cite{Pilaftsis:2003gt} and incorporates a modular $S_4$ flavor structure, which ties the CP phases of the visible and dark sectors to a common complex modulus $\tau$~\cite{Feruglio:2017rjo}.

\subsection{Co-genesis framework}

The two heavy triplet states $\Sigma_{1,2}$ originate from the $S_4$ doublet $\Sigma$ introduced in Sec.~\ref{sec:Model}.  
A small mass splitting $\Delta M_\Sigma = |M_2-M_1|$ is induced by the modular-symmetry--breaking parameter $M_\epsilon$ in Eq.~(\ref{eq:2.8}), naturally realizing the quasi-degenerate spectrum required for resonant enhancement~\cite{Pilaftsis:2003gt}.

The relevant interactions governing leptogenesis and dark-matter asymmetry generation are
\begin{equation}
\mathcal{L} \supset 
- (Y_{\nu})_{\alpha i}\, \overline{L_\alpha}\, \widetilde{H}_u\, \Sigma_{i}
- (Y_{\chi})_i\, \overline{\chi}\,\phi\, \Sigma_{i}
- \frac{1}{2}\, \overline{\Sigma_{i}^c}\, (M_\Sigma)_{ij}\, \Sigma_{j}
+ \mathrm{h.c.},
\label{eq:Lagrangian}
\end{equation}
where $\widetilde{H}_u=i\sigma_2 H_u^*$.  
  
The modular forms determining these couplings follow from the assignments in Sec.~\ref{sec:Model}:
\[
Y_\nu \propto \alpha_D Y^{(4)}_{\mathbf{3}} + \beta_D Y^{(4)}_{\mathbf{3'}}, \qquad
Y_\chi \propto y_{DM} Y^{(4)}_{\mathbf{2}}, \qquad
M_\Sigma = M Y^{(4)}_{\mathbf{1}} + M_\epsilon Y^{(4)}_{\mathbf{2}}.
\] 
Hence the CP phases in both $Y_\nu$ and $Y_\chi$ originate from the same complex modulus $\tau$.

The heavy fermions $\Sigma_i$ decay through two competing channels,
\begin{equation}
\Sigma_i \to L H_u, \qquad \Sigma_i \to \chi \phi,
\end{equation}
producing CP asymmetries in both the visible and dark sectors.  
Since the dark scalar $\phi$ does not acquire a vev, it acts purely as a final-state field, ensuring the stability of the dark sector.

For the lighter state $\Sigma_1$, the CP asymmetries are defined as
\begin{align}
\epsilon_L &=
\frac{\Gamma(\Sigma_1 \to L H_u) - \Gamma(\Sigma_1 \to \bar{L} H_u^\dagger)}{\Gamma_{\Sigma_1}},
&
\epsilon_\chi &=
\frac{\Gamma(\Sigma_1 \to \chi \phi) - \Gamma(\Sigma_1 \to \bar{\chi} \phi^\dagger)}{\Gamma_{\Sigma_1}},
\label{eq:CPdefs}
\end{align}
where $\Gamma_{\Sigma_1}$ denotes the total decay width of $\Sigma_1$.

\subsection{Resonant CP asymmetries}
\begin{figure*}[t]
\centering
\begin{minipage}{0.32\linewidth}
\centering
\includegraphics[width=1.1\linewidth]{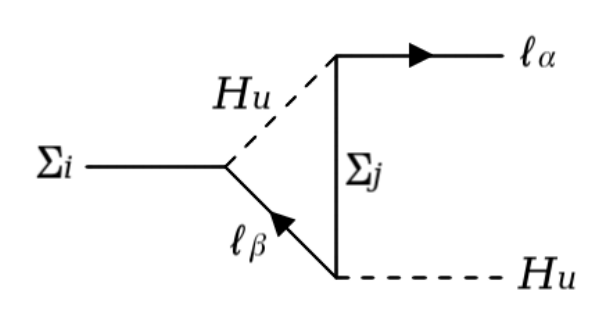}\\(a)
\end{minipage}
\hfill
\begin{minipage}{0.32\linewidth}
\centering
\includegraphics[width=1.1\linewidth]{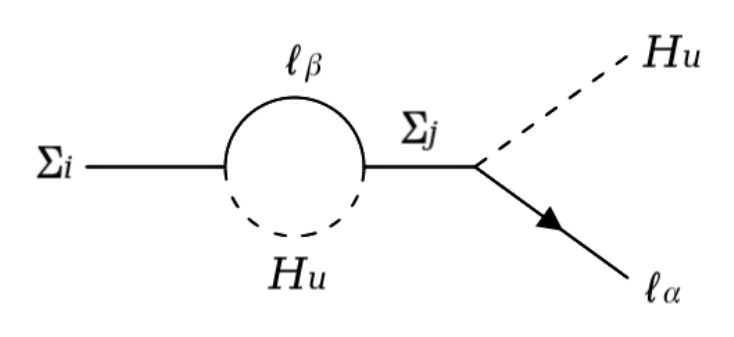}\\(b)
\end{minipage}
\hfill
\begin{minipage}{0.32\linewidth}
\centering
\includegraphics[width=1.1\linewidth]{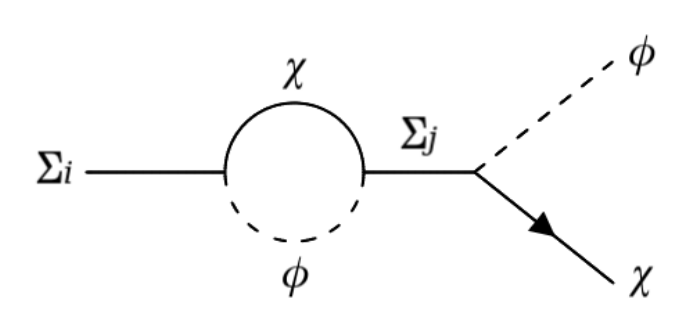}\\(c)
\end{minipage}

\begin{minipage}{0.32\linewidth}
\centering
\includegraphics[width=1.1\linewidth]{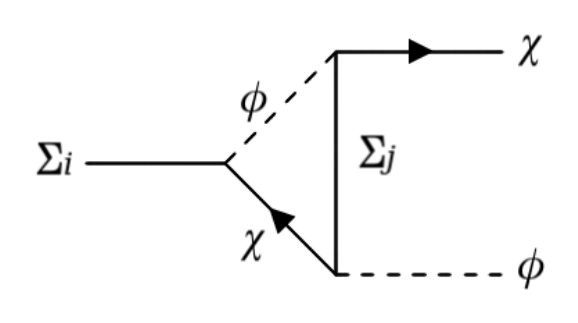}\\(d)
\end{minipage}
\hfill
\begin{minipage}{0.32\linewidth}
\centering
\includegraphics[width=1.1\linewidth]{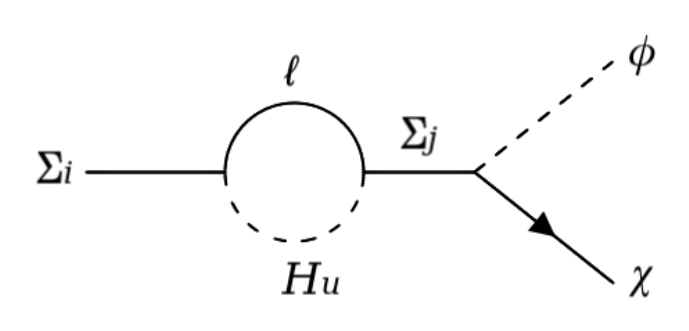}\\(e)
\end{minipage}
\hfill
\begin{minipage}{0.32\linewidth}
\centering
\includegraphics[width=1.1\linewidth]{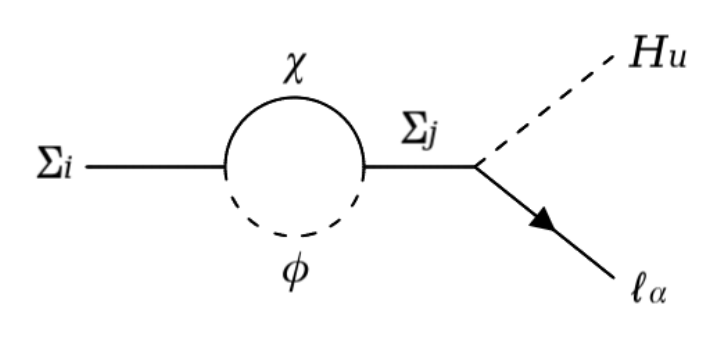}\\(f)
\end{minipage}
\caption{One–loop self–energy diagrams contributing to the CP asymmetries in the visible and dark sectors. 
The interference between tree–level and loop processes with intermediate triplet exchange induces unequal decay rates into 
$\ell_{\alpha} H$ and $\chi \phi$ versus their CP–conjugate channels.}
\label{fig:cp-decay}
\end{figure*}

In the resonant regime, the CP asymmetry is dominated by self-energy diagrams involving the nearly degenerate states $\Sigma_i$ and $\Sigma_j$ ($i\neq j$) see in Fig.~\ref{fig:cp-decay}~\cite{Pilaftsis:2003gt,Dev:2017wwc,Covi:1996wh}.  
Neglecting vertex corrections and keeping only the resonantly enhanced self-energy piece, the CP asymmetries are
\begin{align}
\epsilon_L^i &=
\frac{
\mathrm{Im}\!\left[(Y_\nu^\dagger Y_\nu)_{ij}
\big((Y_\nu^\dagger Y_\nu + 2Y_\chi^\dagger Y_\chi)_{ij}\big)\right]
}{
8\pi\,(Y_\nu^\dagger Y_\nu + Y_\chi^\dagger Y_\chi)_{ii}
}
\;
\frac{(M_i^2 - M_j^2)\,M_i \Gamma_j^{(0)}}
{(M_i^2 - M_j^2)^2 + M_i^2 \Gamma_j^{(0)2}},
\label{eq:epsL}\\[4pt]
\epsilon_\chi^i &=
\frac{
\mathrm{Im}\!\left[(Y_\chi^\dagger Y_\chi)_{ij}
\big((2Y_\nu^\dagger Y_\nu + Y_\chi^\dagger Y_\chi)_{ij}\big)\right]
}{
8\pi\,(Y_\nu^\dagger Y_\nu + Y_\chi^\dagger Y_\chi)_{ii}
}
\;
\frac{(M_i^2 - M_j^2)\,M_i \Gamma_j^{(0)}}
{(M_i^2 - M_j^2)^2 + M_i^2 \Gamma_j^{(0)2}},
\label{eq:epsChi}
\end{align}
where the tree-level width is 
\[
\Gamma_j^{(0)} \simeq 
\frac{M_j}{8\pi}\,
\big[(Y_\nu^\dagger Y_\nu)_{jj} + (Y_\chi^\dagger Y_\chi)_{jj}\big],
\]
including both CP-conjugate decay channels in the standard total-width normalization.

\subsection{Decay widths and branching ratios}

The total decay width of $\Sigma_1$ is
\begin{equation}
\Gamma_{\Sigma_1}
= \Gamma(\Sigma_1 \to L H_u) + \Gamma(\Sigma_1 \to \chi \phi),
\label{eq:totalwidth}
\end{equation}
with partial widths
\begin{equation}
\Gamma(\Sigma_1 \to L H_u)
= \frac{M_{_1}^2}{8\pi v_u^2} \widetilde{m}_1, 
\qquad
\Gamma(\Sigma_1 \to \chi \phi)
= \frac{M_{1}^2}{8\pi v_u^2} \widetilde{m}_{\rm dm},
\label{eq:partialwidths}
\end{equation}
which can be expressed in terms of the effective mass parameters
\begin{equation}
\widetilde{m}_1 = \frac{|Y_\nu|^2 v_u^2}{M_{1}}, 
\qquad
\widetilde{m}_{\rm dm} = \frac{|y_\chi|^2\, v_u^2}{4 M_{_1}}.
\label{eq:mtilde_defs}
\end{equation}
The factor $1/4$ in $\tilde m_{\rm dm}$ arises from the SU(2)$_L$ normalization of the triplet coupling. 
This normalization is consistently included in the total width appearing in the CP-asymmetry denominators.

The branching ratios into visible and dark sectors are therefore
\begin{equation}
{\rm Br}_L = 
\frac{\Gamma(\Sigma_1 \to L H_u)}{\Gamma_{\Sigma_1}}, 
\qquad
{\rm Br}_\chi = 
\frac{\Gamma(\Sigma_1 \to \chi \phi)}{\Gamma_{\Sigma_1}},
\label{eq:branchingratios}
\end{equation}
where the dark Yukawa coupling contributes with an effective weight of $1/4$ in the total width, as reflected in ${\rm Br}_\chi$ through $\tilde m_{\rm dm}$.
\subsection{Boltzmann evolution}

\begin{figure*}[t]
\centering
\begin{minipage}{0.32\linewidth}
\centering
\includegraphics[width=1.1\linewidth]{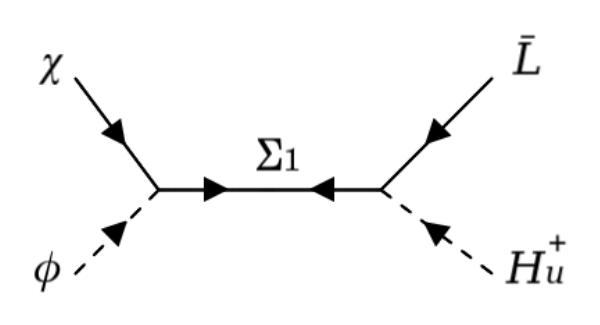}\\
\end{minipage}
\hfill
\begin{minipage}{0.32\linewidth}
\centering
\includegraphics[width=0.8\linewidth]{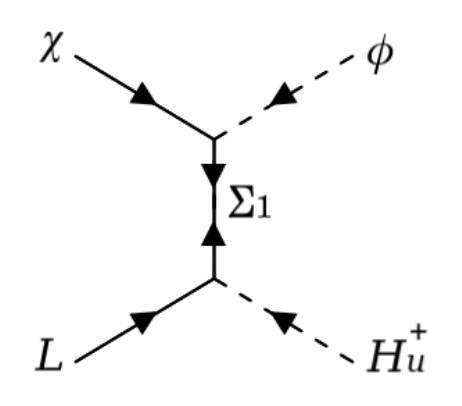}\\
\end{minipage}
\hfill
\begin{minipage}{0.32\linewidth}
\centering
\includegraphics[width=0.8\linewidth]{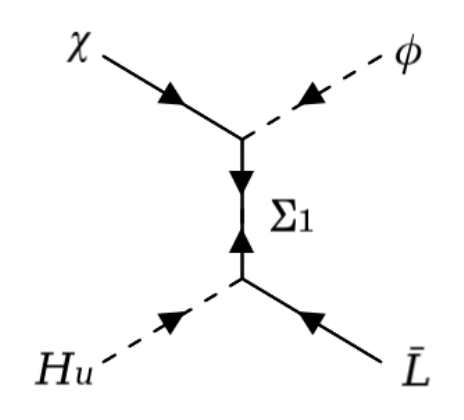}\\
\end{minipage}

\begin{minipage}{0.32\linewidth}
\centering
\includegraphics[width=1.1\linewidth]{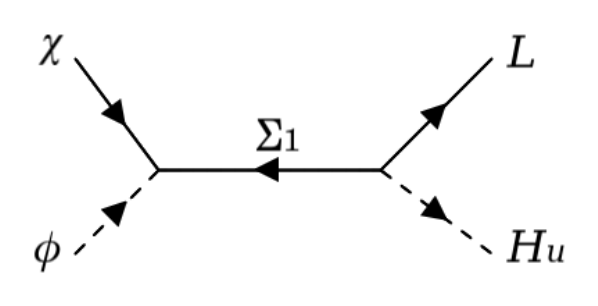}\\
\end{minipage}
\hfill
\begin{minipage}{0.32\linewidth}
\centering
\includegraphics[width=0.8\linewidth]{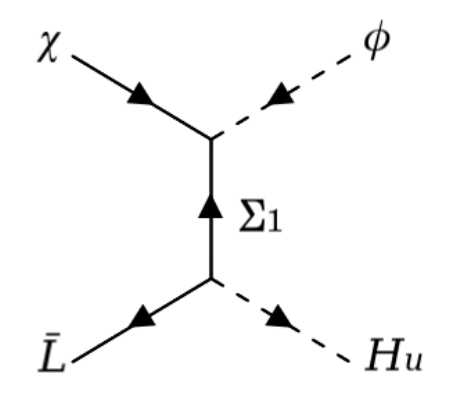}\\
\end{minipage}
\hfill
\begin{minipage}{0.32\linewidth}
\centering
\includegraphics[width=0.8\linewidth]{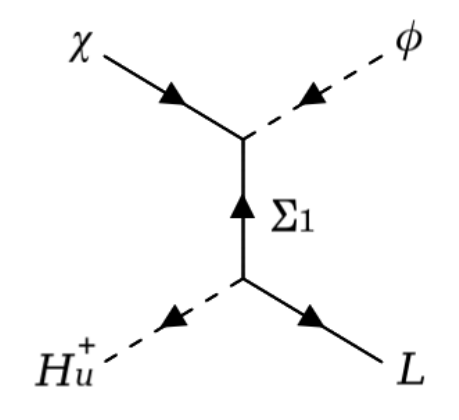}\\
\end{minipage}
\caption{
Feynman diagrams contributing to the $2 \!\to\! 2$ scattering processes 
entering the Boltzmann equations~(\ref{eq:Boltz_L}) and~(\ref{eq:Boltz_DM}). 
The top row shows the $\Delta L = 2$ processes responsible for 
lepton-number-violating washout, while the bottom row displays 
the $\Delta L = 0$ transfer processes that mediate the exchange of the 
asymmetry between the visible and dark sectors. 
}
\label{fig:wash}
\end{figure*}

Defining the washout parameter $K_1 \equiv \Gamma_{\Sigma_1}/H(z=1)$, 
the coupled Boltzmann equations~\cite{Falkowski:2011zr} for the comoving abundances
$Y_{\Sigma_1}=n_{\Sigma_1}/s$, 
$Y_{\Delta L}=Y_L-Y_{\bar L}$, 
and $Y_{\Delta\chi}=Y_\chi-Y_{\bar\chi}$ 
are
\begin{align} \frac{dY_{\Sigma_1}}{dz} &= - \frac{\Gamma_1}{H(z=1)}\, z\, \frac{K_1(z)}{K_2(z)} \bigl(Y_{\Sigma_1} - Y_{\Sigma_1}^{\mathrm{eq}}\bigr), \label{eq:Boltz_Sigma}\\[4pt] 
\frac{dY_{\Delta L}}{dz} &= - \frac{\Gamma_1}{H(z=1)} \Big[ \epsilon_L\, z\, \frac{K_1(z)}{K_2(z)} \bigl(Y_{\Sigma_1}^{\mathrm{eq}} - Y_{\Sigma_1}\bigr) + 2 {\rm Br}_L^2 I_W(z)\, Y_{\Delta L} \notag \\ &\qquad\qquad + {\rm Br}_L {\rm Br}_\chi \Big( I_T^+(z)\, (Y_{\Delta L}+Y_{\Delta \chi}) + I_T^-(z)\, (Y_{\Delta L}-Y_{\Delta \chi}) \Big) \Big], \label{eq:Boltz_L}\\[4pt] \frac{dY_{\Delta \chi}}{dz} &= - \frac{\Gamma_1}{H(z=1)} \Big[ \epsilon_\chi\, z\, \frac{K_1(z)}{K_2(z)} \bigl(Y_{\Sigma_1}^{\mathrm{eq}} - Y_{\Sigma_1}\bigr) + 2 {\rm Br}_\chi^2 I_W(z)\, 
Y_{\Delta \chi} \notag \\ &\qquad\qquad + {\rm Br}_L {\rm Br}_\chi \Big( I_T^+(z)\, (Y_{\Delta L}+Y_{\Delta \chi}) - I_T^-(z)\, (Y_{\Delta L}-Y_{\Delta \chi}) \Big) \Big], \label{eq:Boltz_DM} \end{align}
where $H(z)$ is the Hubble rate in the radiation-dominated era and $K_{1,2}(z)$ are modified Bessel functions.  
The functions $I_W(z)$ and $I_T^\pm(z)$ encode thermally averaged washout and transfer contributions from inverse decays and off-shell $\Sigma_1$-mediated scatterings~\cite{Falkowski:2011zr,Mahapatra:2024avc}, as detailed in Appendix~C.
These equations capture the interplay among decay, washout, and transfer processes that shape both asymmetries, as illustrated in Fig.~\ref{fig:wash}.

\subsection{Relic asymmetries and dark matter abundance}
After $\Sigma_1$ decays freeze out, $Y_{\Delta L}$ and $Y_{\Delta\chi}$ become conserved.  
The lepton asymmetry is partially converted into a baryon asymmetry by electroweak sphalerons,
\begin{equation}
Y_B = c_{\rm sph}\, Y_{\Delta L}, 
\qquad
c_{\rm sph} = \frac{8}{23},
\label{eq:sphaleron}
\end{equation}
which assumes Standard Model field content and efficient sphaleron conversion above the electroweak scale~\cite{Buchmuller:2005eh}.
The symmetric component of $\chi$ annihilates efficiently through $t$-channel 
$\phi$ exchange {since $m_\chi \simeq 0.1{-}2\,{\rm GeV} \ll m_\phi \simeq 100\,{\rm GeV}$, 
rendering $\chi\bar{\chi}\to\phi\phi$ kinematically forbidden}, leaving only 
the asymmetric relic contribution relevant at late times. The present-day energy-density ratio of dark matter to baryons is
\begin{equation}
\frac{\Omega_\chi}{\Omega_B}
= \frac{m_\chi}{m_p}\,
  \frac{|Y_{\Delta \chi}|}{|Y_{\Delta L}|},
\label{eq:Omega_ratio}
\end{equation}
where $m_p$ is the proton mass.  
The observed ratio $\Omega_\chi/\Omega_B \simeq 5.4$~\cite{Planck:2018vyg} implies that comparable visible and dark asymmetries naturally correspond to $m_\chi  \simeq 0.1{-}2$ GeV.

Once $\Sigma_1$ has decayed completely, the asymmetries saturate to constant values $Y_{\Delta L}^\infty$ and $Y_{\Delta\chi}^\infty$, leading to~\cite{Zurek:2013wia}
\begin{equation}
\Omega_\chi h^2 
= \frac{m_\chi\, s_0\, |Y_{\Delta \chi}^\infty|}{\rho_c/h^2},
\qquad
\rho_c/h^2 = 1.05\times10^{-5}~\mathrm{GeV\,cm^{-3}},
\label{eq:Omegachi}
\end{equation}
and therefore
\begin{equation}
m_\chi 
\simeq 
m_p\, \frac{\Omega_\chi}{\Omega_B}\,
\left|\frac{Y_{\Delta L}^\infty}{Y_{\Delta \chi}^\infty}\right|
\simeq 
5.4\, m_p\,
\left|\frac{Y_{\Delta L}^\infty}{Y_{\Delta \chi}^\infty}\right|.
\label{eq:mchi_relation}
\end{equation}
If ${\rm Br}_\chi > {\rm Br}_L$, the dark asymmetry is larger and Eq.~\eqref{eq:mchi_relation} predicts a correspondingly heavier ADM candidate, linking the DM and neutrino mass scales through the same modular origin of CP violation.

\noindent
Overall, this framework quantitatively connects neutrino masses, baryon asymmetry, and dark matter abundance, all governed by the common modular parameter $\tau$ that controls both flavor and CP violation.

\section{Numerical Analysis}
\label{sec:num}

\subsection{Neutrino Sector}
\begin{table}[htbp]
\centering
\caption{NuFIT~5.2 (2024) best-fit values and $3\sigma$ ranges for normal hierarchy.}
\renewcommand{\arraystretch}{1.2}
\setlength{\tabcolsep}{12pt}
\begin{tabular}{c c c} 
\hline
\textbf{Parameter} & \textbf{Best-fit $\pm$ 1$\sigma$} & \textbf{3$\sigma$ range}  \\ 
\hline
$\sin^2 \theta_{12}$ & $0.307^{+0.012}_{-0.011}$ & $0.275 - 0.345$  \\  
$\sin^2 \theta_{23}$ & $0.561^{+0.012}_{-0.015}$ & $0.430 - 0.596$  \\  
$\sin^2 \theta_{13}$ & $0.02195^{+0.00054}_{-0.00058}$ & $0.02023 - 0.02376$  \\
$\Delta m^2_{21}\ [10^{-5}\ \mathrm{eV}^2]$ & $7.49^{+0.19}_{-0.19}$ & $6.92 - 8.05$  \\  
$\Delta m^2_{31}\ [10^{-3}\ \mathrm{eV}^2]$ & $2.534^{+0.025}_{-0.023}$ & $2.463 - 2.606$  \\
\hline
\end{tabular}
\label{tab:3}
\end{table}
The numerical analysis has been performed for the normal hierarchy of neutrino masses within the modular $S_4$ framework. 
The light-neutrino mass matrix is obtained from the type-III seesaw relation 
$M_\nu = -M_D M_\Sigma^{-1} M_D^{T}$~\cite{Foot:1988qv,Ma:1998dn}, 
where all entries of $M_D$ and $M_\Sigma$ are determined by the modular forms $Y_i(\tau)$ defined in Appendix A. 
The physical lepton mixing matrix is constructed as $U_{\rm PMNS} = U_\ell^\dagger U_\nu$~\cite{PDG2024}, with $U_\ell$ and $U_\nu$ diagonalizing $M_\ell M_\ell^\dagger$ and $M_\nu$, respectively. 
All oscillation observables are computed in the standard PMNS parametrization,\footnotemark and the resulting predictions are compared with the NuFIT~5.2~(2024) global analysis for normal ordering~\cite{Esteban:2024vvu}. 
\footnotetext{Explicit definitions of $\sin^2\theta_{ij}$, $\delta_{\rm CP}$, and $m_{\beta\beta}$ are given in Appendix~B.} 
The corresponding $3\sigma$ ranges are summarized in Table~\ref{tab:3}.

\begin{figure}[t]
\centering
\begin{minipage}{0.95\textwidth}
\centering
\includegraphics[width=0.49\textwidth]{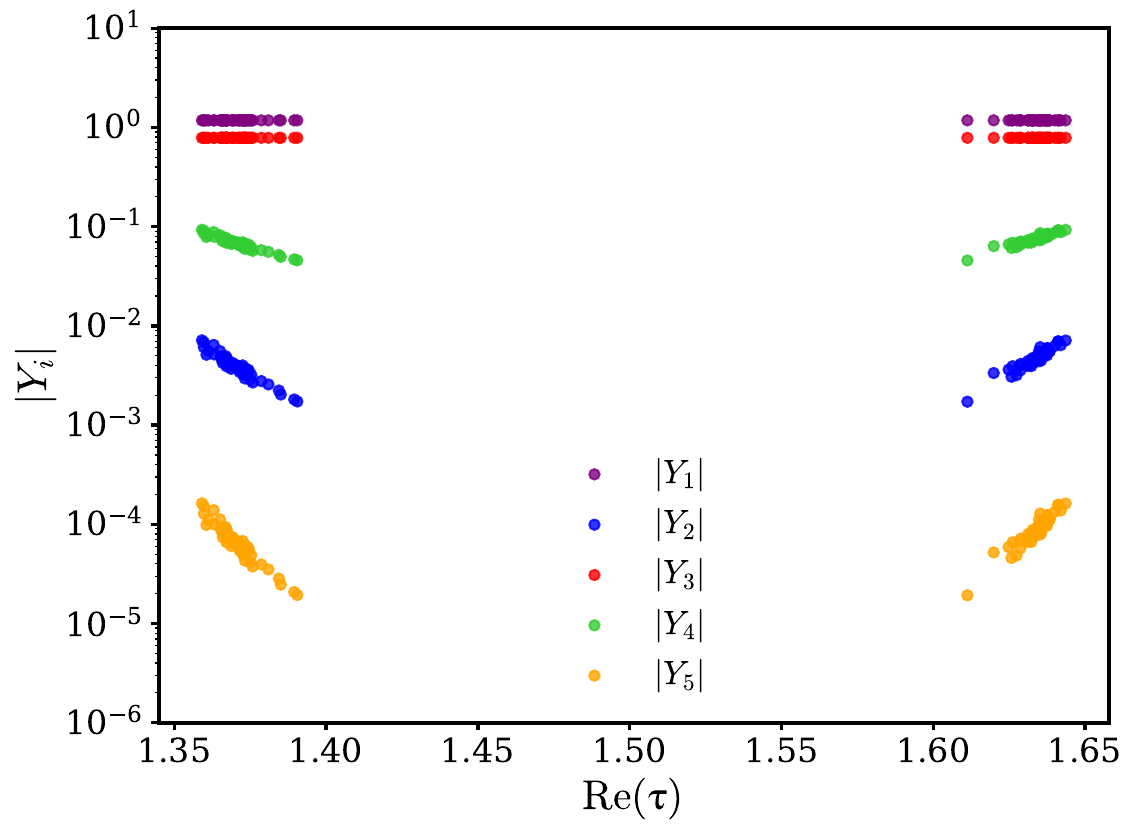}
\includegraphics[width=0.49\textwidth]{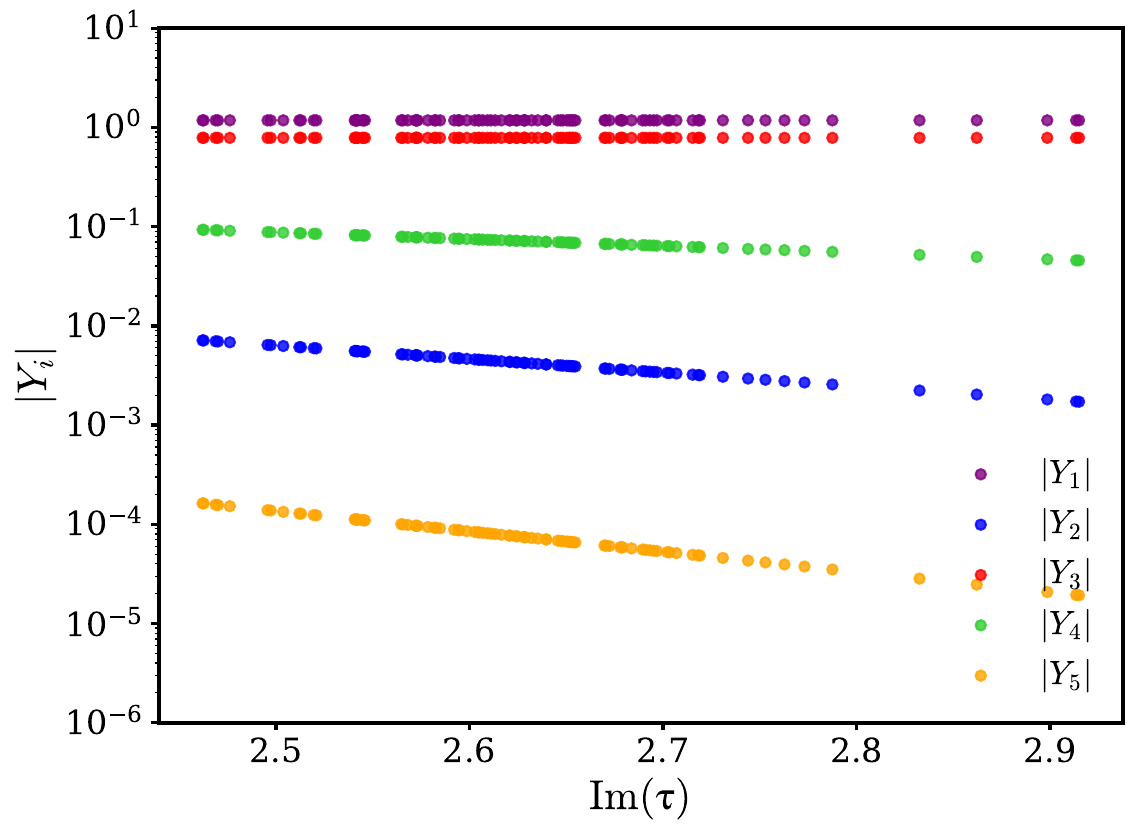}
\end{minipage}

\vspace{0.3cm}

\begin{minipage}{0.95\textwidth}
\centering
\includegraphics[width=0.5\textwidth]{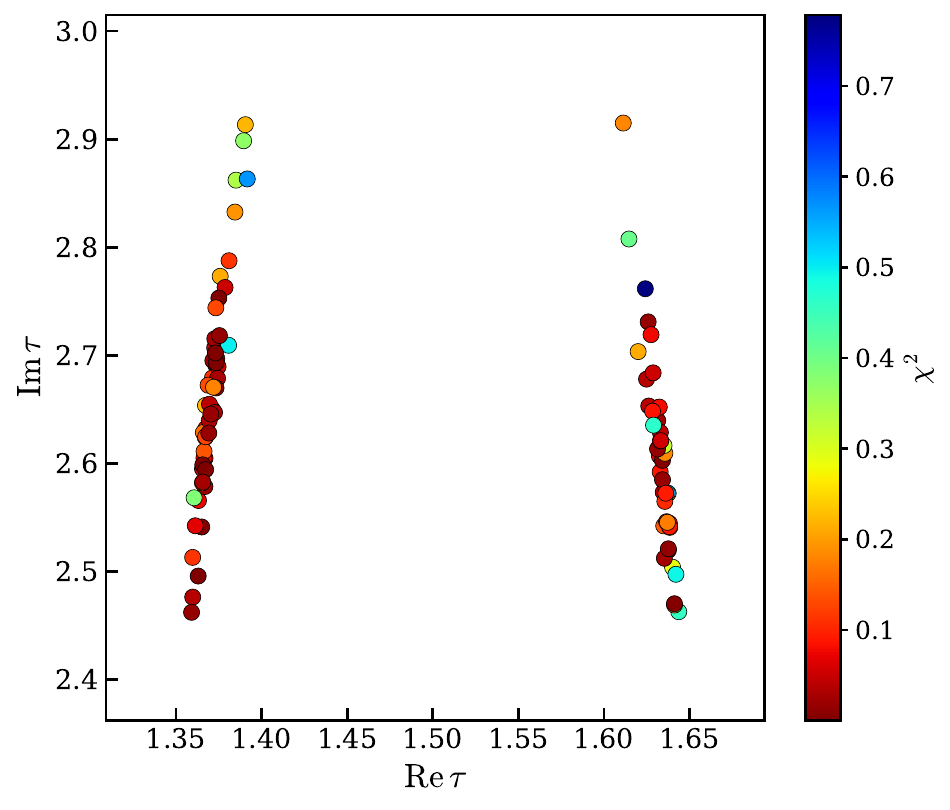}
\end{minipage}

\caption{
Allowed region of the complex modulus $\tau$ in the modular $S_4$ framework.
Upper: absolute values of modular Yukawa forms $|Y_i|$ versus $\operatorname{Re}\tau$ (left) and $\operatorname{Im}\tau$ (right).
Lower: $\chi^2$ distribution in the $(\operatorname{Re}\tau, \operatorname{Im}\tau)$ plane.
All displayed points lie within the NuFIT~5.2 $3\sigma$ ranges.
}
\label{fig:tau_plane}
\end{figure}

The input parameters are varied within the domains
\[
\begin{array}{c@{\qquad}c@{\qquad}c}
1.0 < \mathrm{Re}\,\tau < 2.0, & 0.5 < \mathrm{Im}\,\tau < 3.0, & 0.5 < \alpha < 1.6, \\[2pt]
0.3 < \beta < 1.1, & 1.0 < \gamma < 1.6, & 10^{-3} < \alpha_D < 10^{-2}, \\[2pt]
10^{-5} < \beta_D < 10^{-3}, & 5\times 10^{2}~\mathrm{GeV} < M < 5\times 10^{3}~\mathrm{GeV}, & 10^{5}~\mathrm{GeV} < M_\epsilon < 5\times 10^{7}~\mathrm{GeV}.
\end{array}
\]
\begin{figure}[t]
\centering
\begin{minipage}{0.95\textwidth}
\centering
\includegraphics[width=0.48\textwidth]{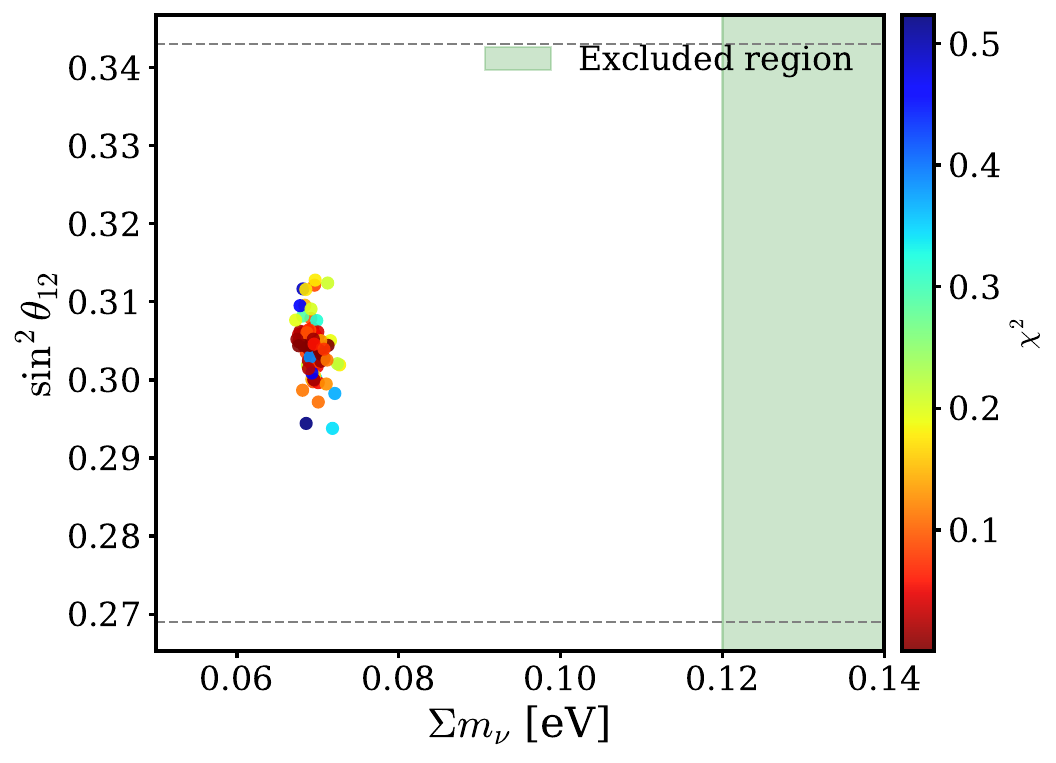}
\includegraphics[width=0.48\textwidth]{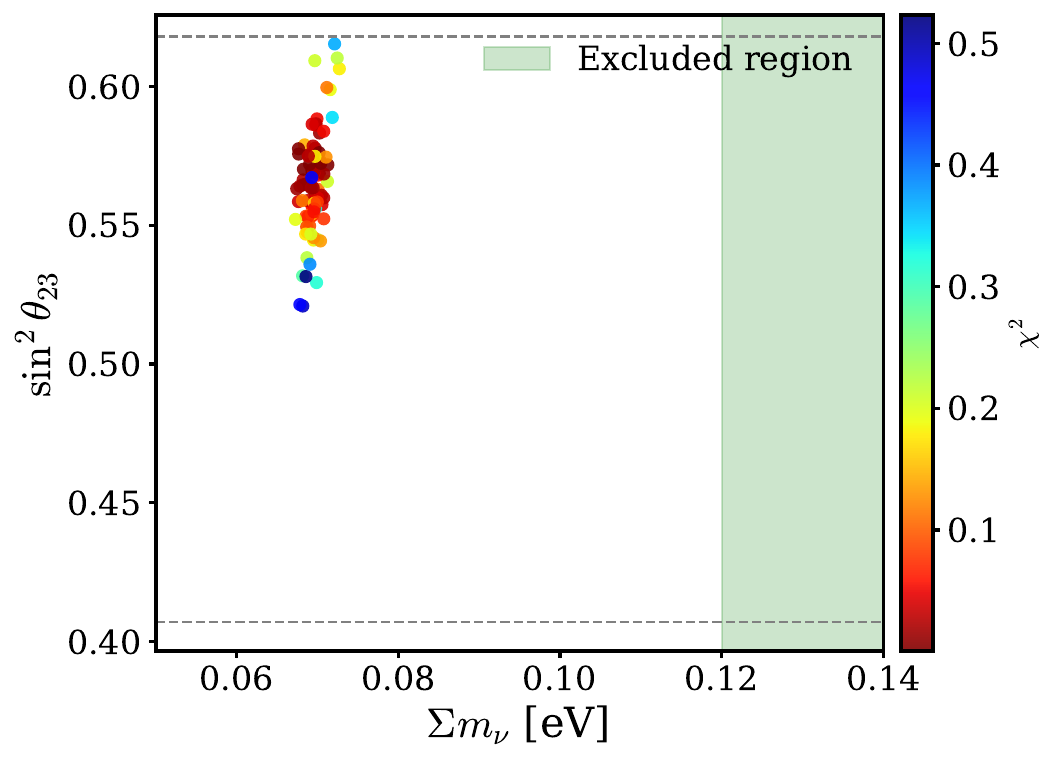}
\end{minipage}

\vspace{0.3cm}

\begin{minipage}{0.95\textwidth}
\centering
\includegraphics[width=0.48\textwidth]{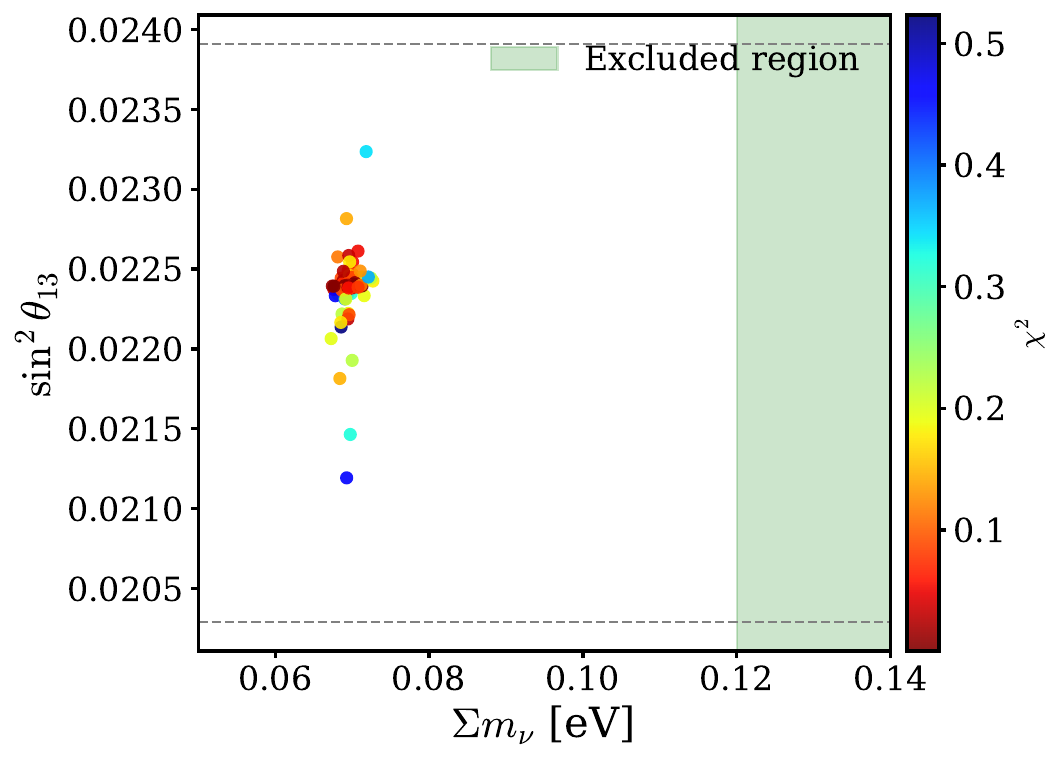}
\end{minipage}

\caption{
Correlations between the total neutrino mass $\sum m_\nu$ and the mixing angles $\sin^2\theta_{12}$ (upper left), $\sin^2\theta_{23}$ (upper right), and $\sin^2\theta_{13}$ (lower panel).
Dashed lines denote the $3\sigma$ ranges from NuFIT~5.2, and the shaded band shows the cosmological bound $\sum m_\nu < 0.12~\mathrm{eV}$.
}
\label{fig:mixing_sum}
\end{figure}

For each parameter set, the modular Yukawa couplings are determined by the chosen $\tau$, thereby fixing the flavor structure entirely through the $S_4$ modular forms~\cite{Feruglio:2017rjo}. 
The resulting parameter space is constrained by the NuFIT~5.2 $3\sigma$ limits on mixing angles and mass-squared differences~\cite{Esteban:2024vvu}, as well as the Planck bound $\sum m_\nu < 0.12~\mathrm{eV}$~\cite{Planck:2018vyg}.

The region of the complex modulus $\tau$ consistent with neutrino oscillation data is displayed in
Fig.~\ref{fig:tau_plane}, showing that acceptable solutions cluster around
two narrow bands near $\mathrm{Re}\,\tau\simeq1.35$ and
$\mathrm{Re}\,\tau\simeq1.63$ with $\mathrm{Im}\,\tau\simeq2.6$--$2.8$.
These regions correspond to points where the modular Yukawa structures
reproduce all neutrino mixing angles and mass-squared differences within
the $3\sigma$ NuFIT~5.2 ranges, thereby phenomenologically fixing
the modulus.  In this framework, $\tau$ is not stabilized by an explicit
potential but is effectively determined by requiring simultaneous
agreement with low-energy neutrino observables, a standard approach in
modular-flavor analyses
\cite{Ding:2019gof, Novichkov:2020uzt}. We have verified that restricting the scan 
to the fundamental domain 
$|\mathrm{Re}\,\tau| \leq 1/2$ yields 
equivalent viable solutions at 
$\mathrm{Re}\,\tau \simeq \pm(0.37$--$0.40)$ 
with $\mathrm{Im}\,\tau \simeq 2.6$--$2.9$, 
confirming that the solutions reported in 
Fig.~3 are the modular $T$-images of 
solutions within the fundamental domain. The localization of $\tau$
close to the self-dual point $\omega=e^{2\pi i/3}$ reflects the geometric
properties of modular forms, which naturally yield hierarchical and stable
Yukawa textures in this region.  The magnitudes of the modular couplings
exhibit the ordering $|Y_1|\gg|Y_3|>|Y_2|>|Y_4|>|Y_5|$ as shown in upper panel of Fig.~\ref{fig:tau_plane}, consistent with
realistic lepton mass hierarchies.  The apparent flatness of the modular
Yukawas arises from the exponential suppression of higher-order
$q$-terms ($q=e^{2\pi i\tau}$) for large $\operatorname{Im}\tau$, leading
to quasi-degenerate modular textures in a phenomenologically preferred
$\tau$ region that simultaneously accounts for flavor and CP observables.

The correlations between $\sum m_\nu$ and the mixing angles are shown in Fig.~\ref{fig:mixing_sum}. 
All viable points lie below the cosmological limit $\sum m_\nu < 0.12~\mathrm{eV}$ and fall within the $3\sigma$ regions of the global fit. 
The results indicate $\sum m_\nu \simeq (0.06$–$0.08)\,\mathrm{eV}$, $\sin^2\theta_{12}\simeq 0.30$, $\sin^2\theta_{23}\simeq 0.57$, and $\sin^2\theta_{13}\simeq 0.022$, providing an excellent match to data and confirming that the model favors the normal ordering.

\begin{figure}[t]
\centering
\includegraphics[width=0.51\textwidth]{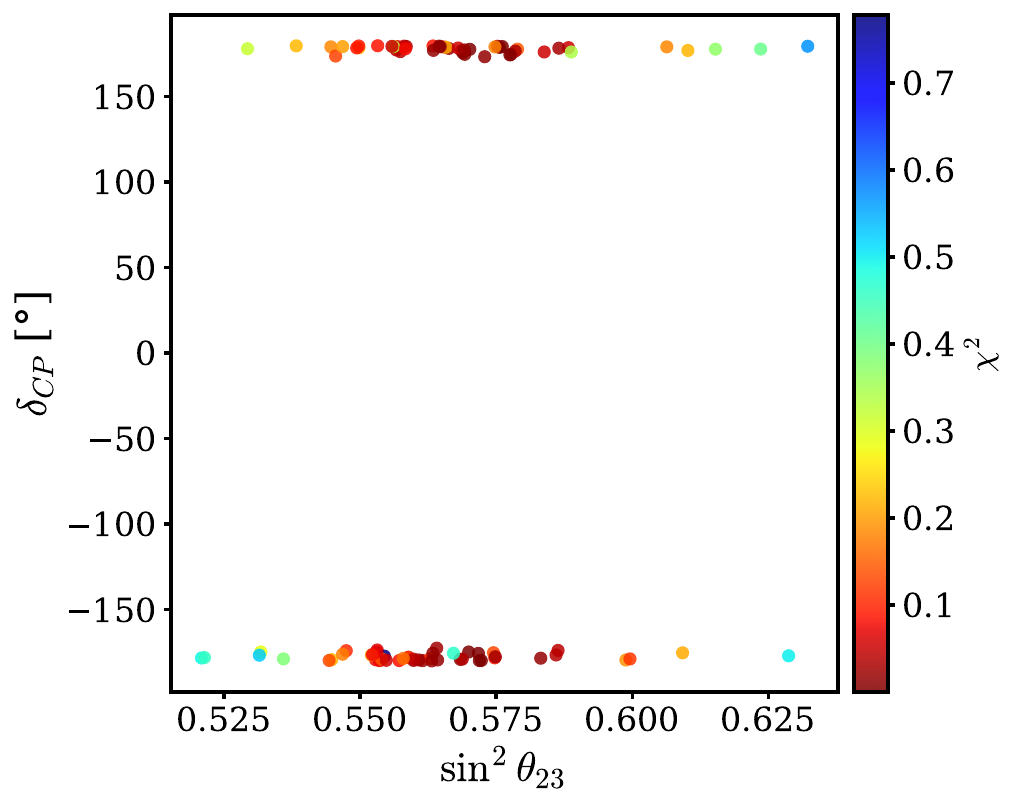}
\includegraphics[width=0.48\textwidth]{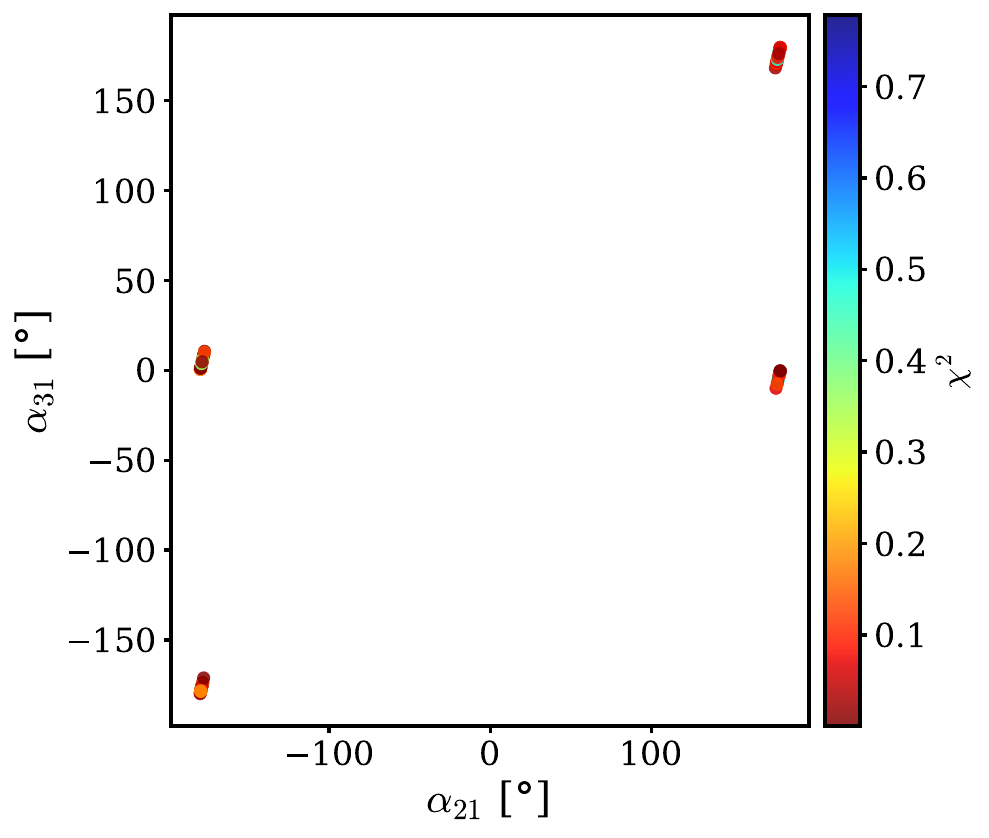}
\caption{
Left: correlation between $\delta_{\rm CP}$ and $\sin^2\theta_{23}$, showing near-maximal CP violation.
Right: predicted Majorana phase structure $(\alpha_{21},\alpha_{31})$ exhibiting discrete clustering near $(0,\pi)$ and $(\pm\pi/2)$, a distinctive imprint of the modular symmetry.
}
\label{fig:delta_majorana}
\end{figure}

The atmospheric mixing angle and CP phase satisfy $\sin^2\theta_{23}\simeq0.56$–$0.59$ and $\delta_{\rm CP}\in[\pm150^\circ,\pm180^\circ]$, as displayed in Fig.~\ref{fig:delta_majorana}, implying nearly maximal CP violation. 
The predicted Majorana phases cluster around $(0,\pi)$ and $(\pm\pi/2)$, reflecting the discrete modular structure of the Yukawa sector.

\begin{figure}[t]
\centering
\includegraphics[width=0.60\textwidth]{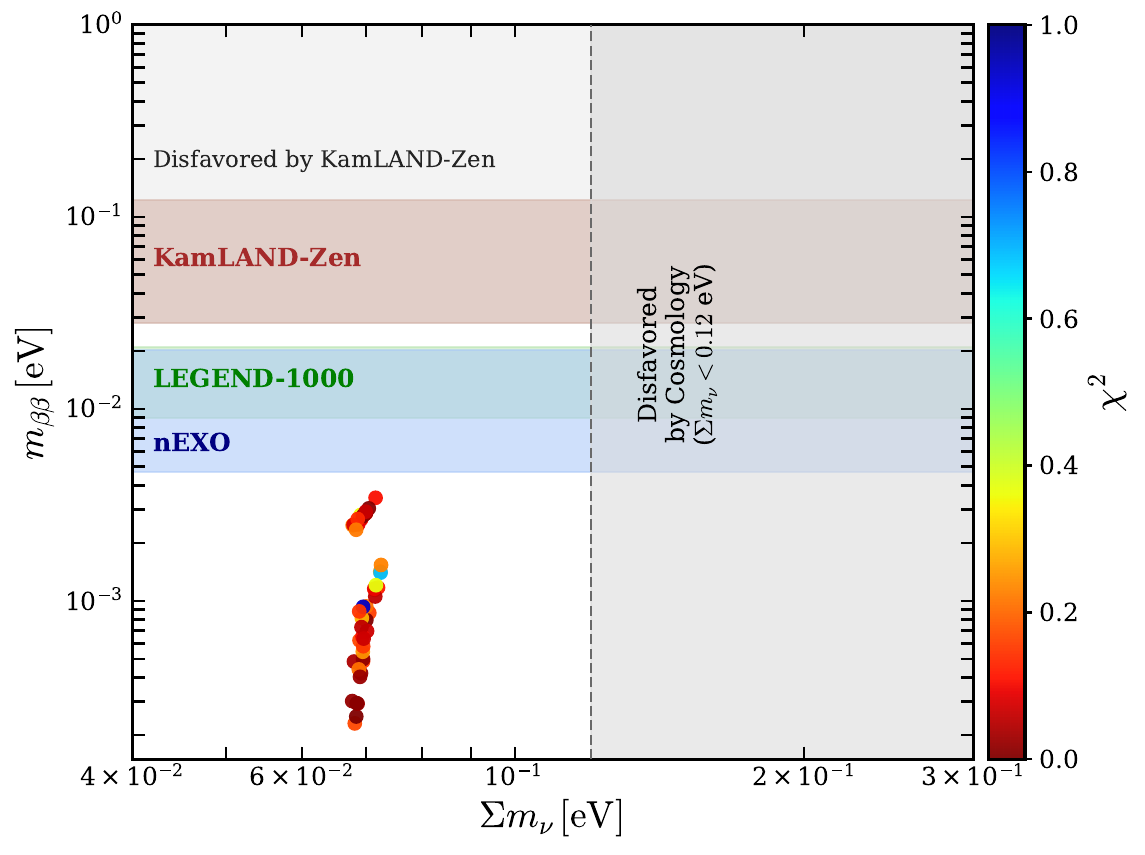}
\caption{
Effective Majorana mass $m_{\beta\beta}$ versus the sum of neutrino mass $\sum m_\nu$.
Colored bands denote the current and projected sensitivities of KamLAND-Zen, LEGEND-1000, and nEXO.
}
\label{fig:mbb_m1}
\end{figure}

The effective Majorana mass relevant for neutrinoless double beta decay, 
$m_{\beta\beta}=|m_1 U_{e1}^2 + m_2 U_{e2}^2 + m_3 U_{e3}^2|$~\cite{Abhishek:2025ijv}, 
is found to be in the range 
$m_{\beta\beta}\simeq(8$–$18)\times10^{-3}\,\mathrm{eV}$, 
as shown in Fig.~\ref{fig:mbb_m1}. 
This lies below the current KamLAND-Zen bound~\cite{Abe:2024wvc} but within the future reach of nEXO~\cite{Adhikari_2021} and LEGEND-1000~\cite{collaboration2021legend1000}.

The modular $S_4$ framework with type-III seesaw successfully reproduces all neutrino oscillation observables within $3\sigma$ without fine-tuning, with the entire flavor structure governed by a single modulus $\tau$.
Since the same Yukawa couplings $(Y_\nu)$ and triplet masses $(M_1, M_2)$ also control the decay asymmetries of the heavy fermions, these neutrino-determined parameters directly constrain the CP-violating sources for leptogenesis.
The next section explores how this common structure naturally links baryon asymmetry and dark matter relic abundance through resonant co-genesis.

\subsection{Baryon–Dark Matter Co-genesis}
Having established the theoretical setup in Section \ref{sec:leptogenesis}, we now present a detailed numerical study of the correlated generation of the baryon asymmetry and ADM within the neutrino-viable modular $S_4$ framework. 
All Yukawa couplings, CP phases, and decay parameters entering leptogenesis are derived from the modular forms $Y_i(\tau)$ fixed by the same complex modulus $\tau$ that governs the neutrino sector. 
Hence, no additional free phases are introduced for baryogenesis or ADM production~\cite{Novichkov:2020uzt}.

The CP asymmetries $\epsilon_L$ and $\epsilon_\chi$ are evaluated using the modular Yukawa structures defined in Eqs.~(\ref{eq:epsL})–(\ref{eq:epsChi}), within the neutrino-consistent parameter space obtained in Section~\ref{sec:num}. 
The phenomenologically relevant region of the complex modulus~$\tau$ is restricted to two narrow bands centered at $\mathrm{Re}\,\tau \simeq 1.37$ and $1.64$ with $\mathrm{Im}\,\tau \simeq 2.6\text{--}2.8$, where the modular forms develop sizable imaginary components. For each allowed $\tau$, the corresponding Yukawa matrices $(Y_\nu, Y_\chi)$ are constructed and inserted into the loop-induced decay asymmetry expressions, yielding the visible and dark sector CP parameters $\epsilon_L$ and $\epsilon_\chi$. The relative magnitudes of $\epsilon_L$ and $\epsilon_\chi$ are further controlled by the
portal coupling $y_{\mathrm{DM}}$ appearing in Eqs.~(\ref{eq:2.9}).
Although the modular structure fixes the relative phases of the two sectors,
the value of $y_{\mathrm{DM}}$ determines their relative decay branching fractions
and hence the dominant source of CP violation.
For $y_{\mathrm{DM}} \lesssim 10^{-3}$, the dark decay
$\Sigma_1\!\to\!\chi\phi$ dominates, yielding a larger dark asymmetry.
At $y_{\mathrm{DM}}\!\sim\!10^{-2}$, the visible and dark CP asymmetries
become nearly identical, $|\epsilon_L|\!\simeq\!|\epsilon_\chi|$,
reflecting their common modular origin. 
This regime corresponds to balanced co-genesis, where both sectors
freeze out with comparable asymmetries. A representative illustration of the $y_{\mathrm{DM}}$ dependence of the CP asymmetry ratio is provided in Appendix D, confirming the robustness of our numerical results.

\begin{figure}[htbp]
\centering
\begin{minipage}{0.95\textwidth}
\centering
\includegraphics[width=0.48\textwidth]{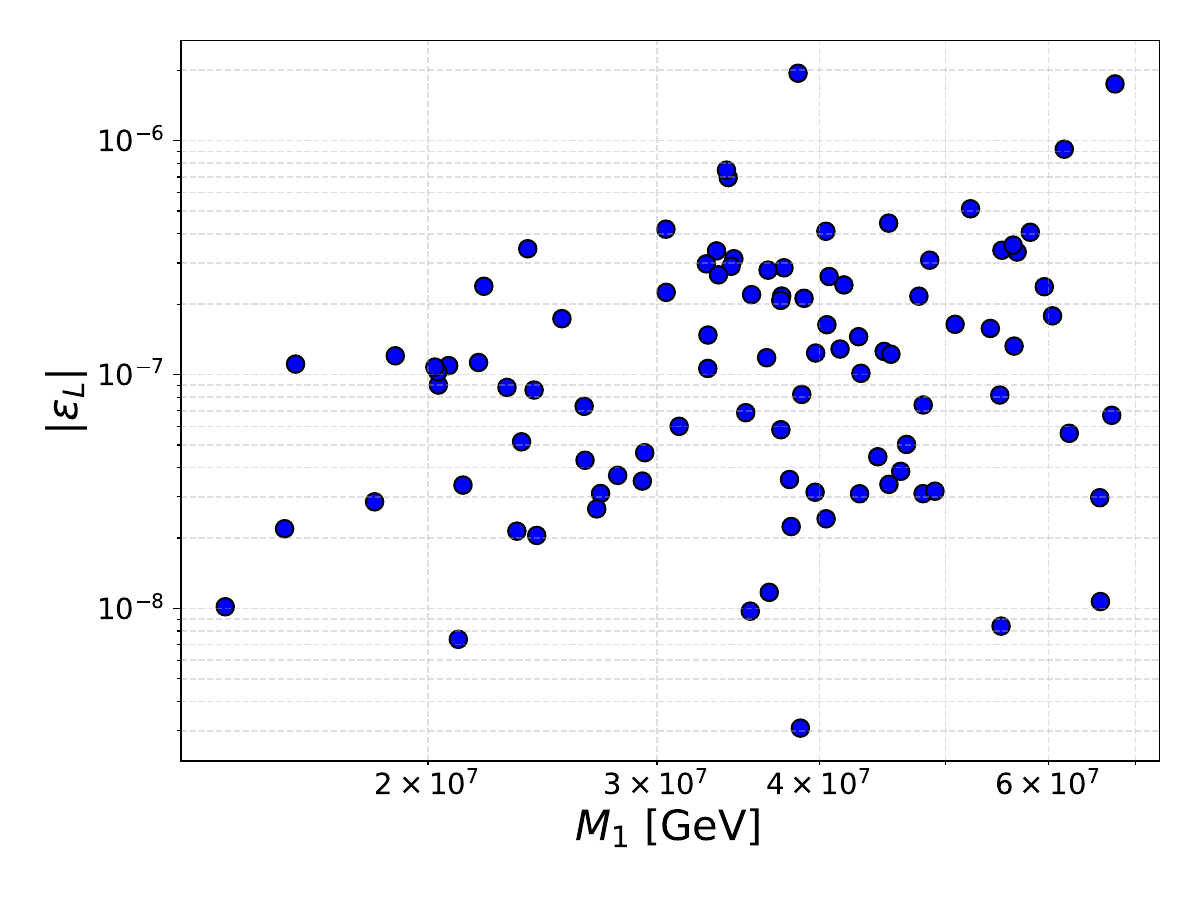}
\includegraphics[width=0.48\textwidth]{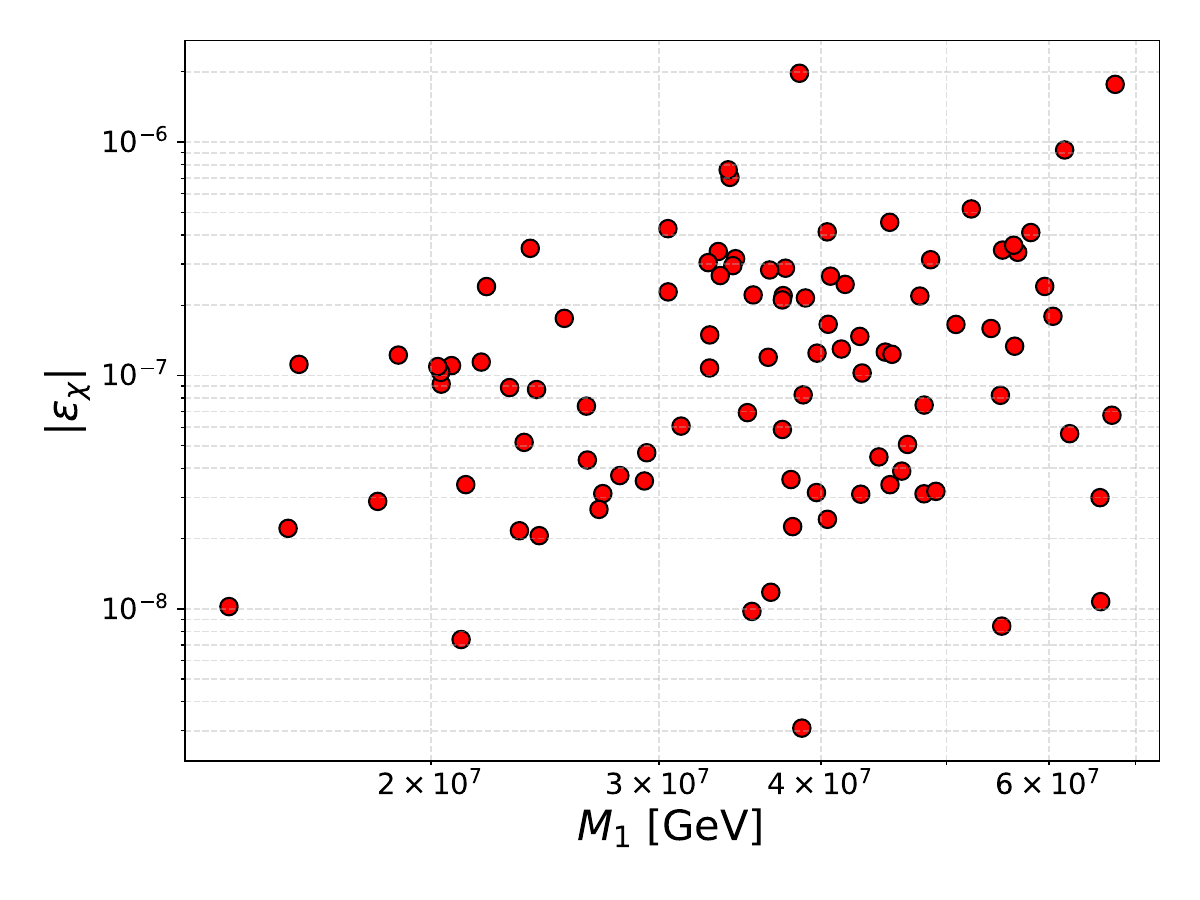}
\includegraphics[width=0.48\textwidth]{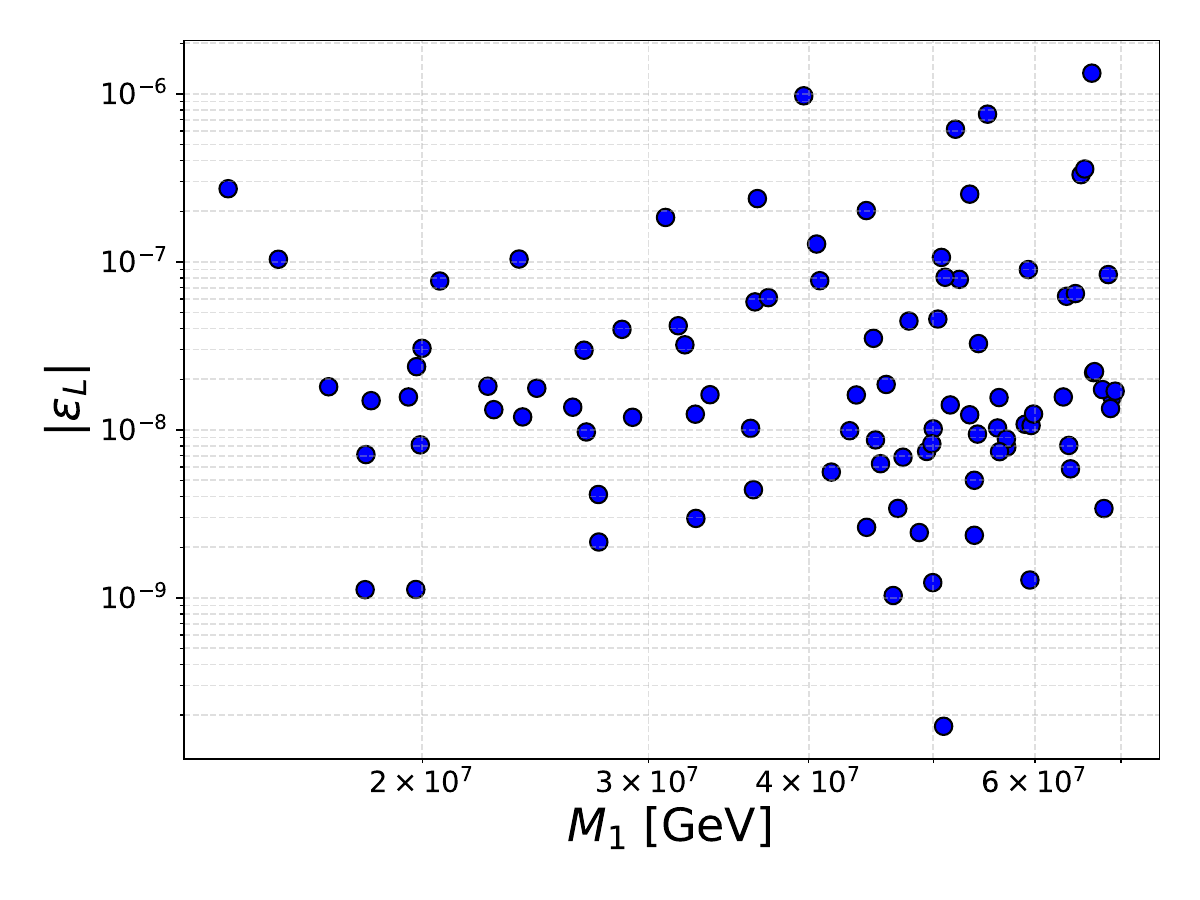}
\includegraphics[width=0.48\textwidth]{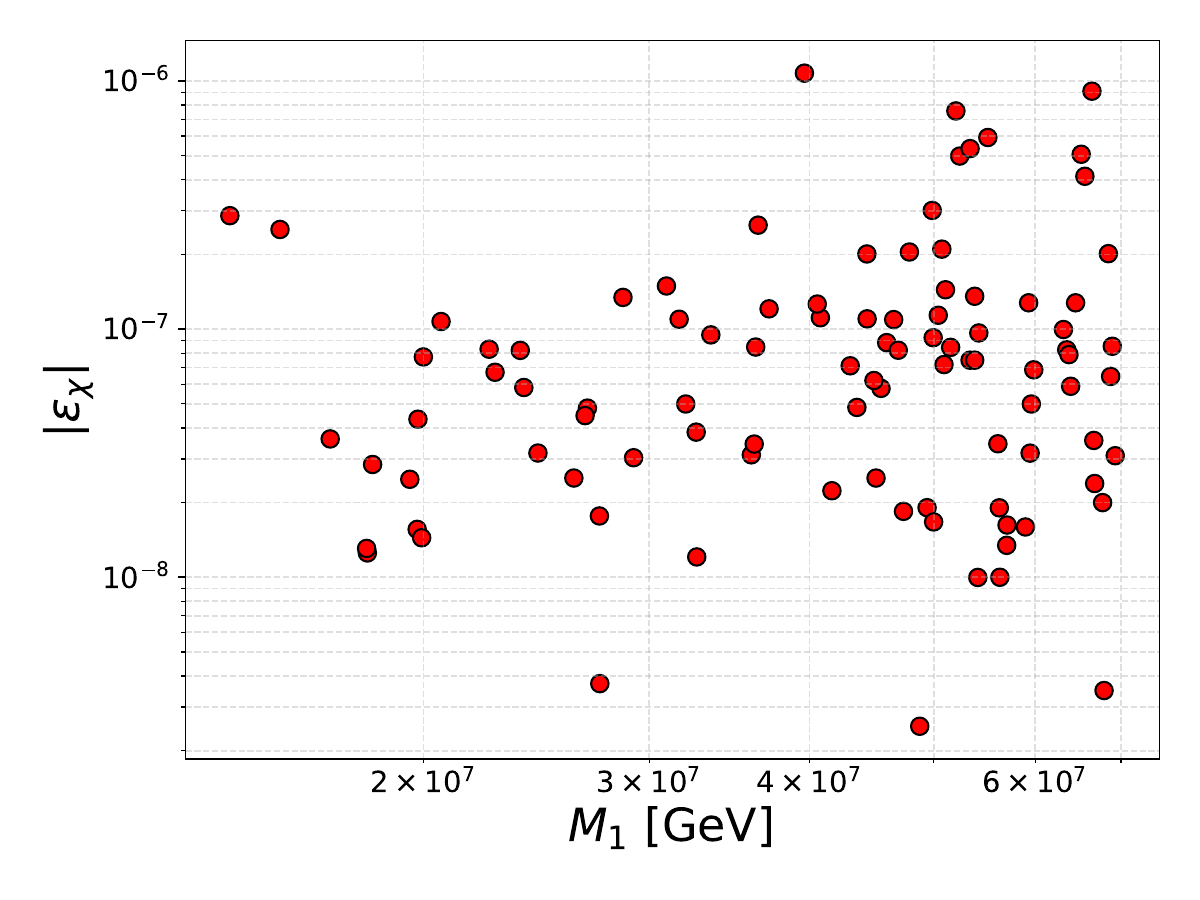}
\end{minipage}
\caption{
CP asymmetries as functions of the lightest heavy-triplet mass $M_1$ 
for two representative values of the dark-sector Yukawa coupling $y_{\mathrm{DM}}$.
Upper panels: $y_{\mathrm{DM}} = 10^{-2}$ (balanced co-genesis regime with
$|\epsilon_L| \simeq |\epsilon_\chi|$);
lower panels: $y_{\mathrm{DM}} = 10^{-3}$ (dark-dominated regime with
$|\epsilon_\chi| > |\epsilon_L|$).
Left: visible-sector asymmetry $|\epsilon_L|$; 
right: dark-sector asymmetry $|\epsilon_\chi|$.
}

\label{fig:epsL_M1}
\end{figure}

\begin{figure}[htbp]
\centering
\includegraphics[width=0.55\textwidth]{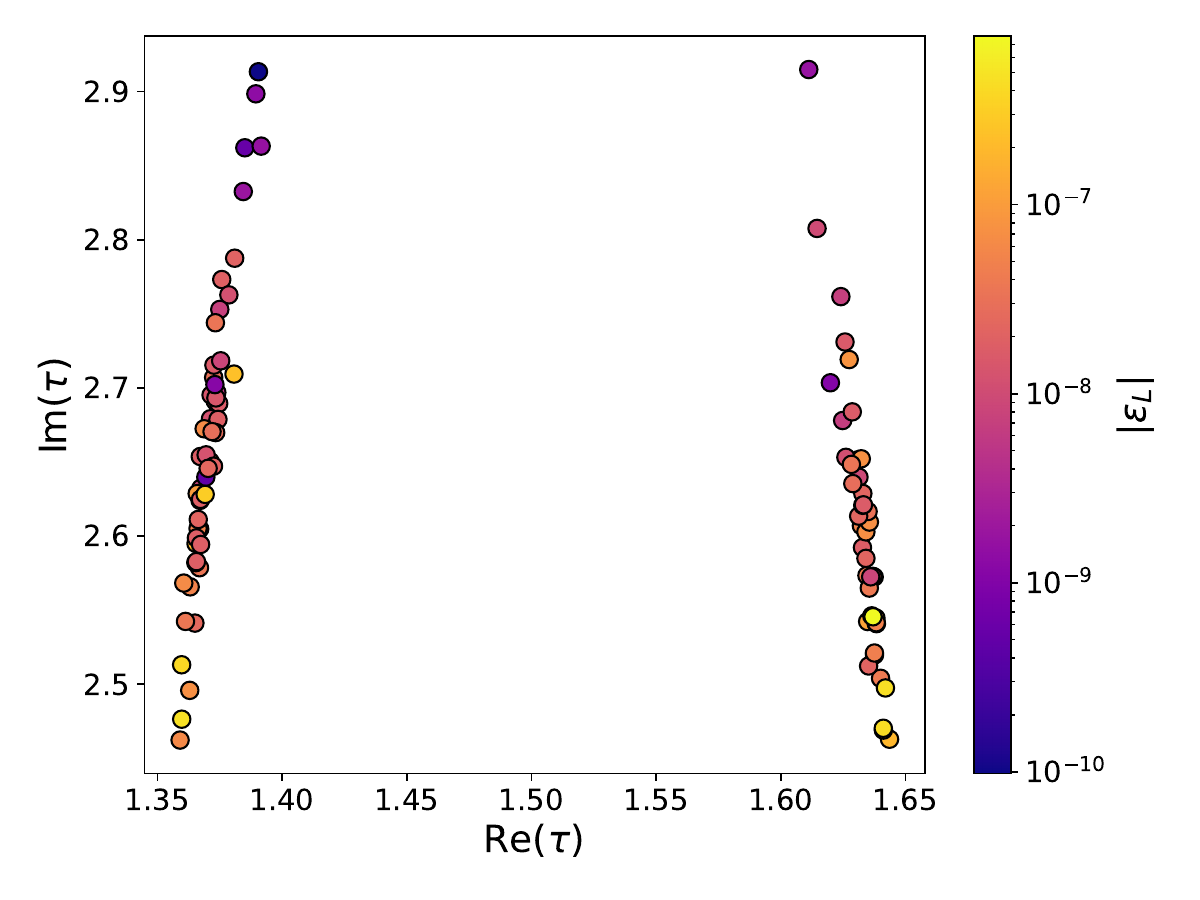}
\caption{
Magnitude of the CP asymmetry $|\epsilon_L|$ in the complex $\tau$ plane. 
The two allowed $\tau$ bands around $\mathrm{Re}\,\tau\simeq1.37$ and $1.64$ with $\mathrm{Im}\,\tau\simeq2.6$–$2.8$ coincide with the neutrino-viable region, confirming that the imaginary part of $\tau$ acts as the sole physical source of CP violation.
}
\label{fig:epsL_tauplane}
\end{figure}

The resulting distributions of $|\epsilon_L|$ and $|\epsilon_\chi|$ as functions of the heavy-triplet mass $M_1$ are shown in Fig.~\ref{fig:epsL_M1}, while the variation of $|\epsilon_L|$ in the complex $\tau$ plane is displayed in Fig.~\ref{fig:epsL_tauplane}. 
Both quantities lie in the range $10^{-9}$–$10^{-6}$ throughout the neutrino-viable
parameter space, sufficient to account for the observed baryon asymmetry after
including washout effects. We note that the CP asymmetries $\epsilon_L$ and $\epsilon_\chi$, as well as the resulting baryon asymmetry $Y_B$, are computed with their full signs throughout the numerical analysis. Parameter points are retained only when they yield the physically required positive baryon asymmetry, $Y_B > 0$.
The absolute values shown in the figures are used solely for visualization on logarithmic scales. Therefore, imposing the correct sign of $Y_B$ does not introduce any additional constraint beyond the viable parameter space already selected. The mild spread visible in Fig.~\ref{fig:epsL_M1} thus reflects not only the small
modular-phase differences but also the variation in $y_{\mathrm{DM}}$ that controls
the visible–dark balance.

The resonant enhancement characteristic of this regime amplifies the CP violation by a factor of ${\cal O}(10^{2\text{--}3})$, compensating for the relatively small Yukawa couplings $|Y_{\nu,\chi}| \sim 10^{-3}$ without the need for any fine-tuning. Although both $\epsilon_L$ and $\epsilon_\chi$ are governed by the same complex modulus $\tau$, they are not exactly proportional. 
The visible-sector Yukawa matrix $Y_\nu \!\sim\! (Y^{(4)}_{\mathbf{3}}, Y^{(4)}_{\mathbf{3}'})$ and the dark-sector coupling $Y_\chi \!\sim\! Y^{(4)}_{\mathbf{2}}$ transform under different $S_4$ multiplets and carry distinct modular weights. 
As a result, their complex phases are correlated but not identical, leading to comparable—though not perfectly matched—CP asymmetries in the two sectors. 
This feature is evident in Figs.~\ref{fig:epsL_M1}(a,b), where both $|\epsilon_L|$ and $|\epsilon_\chi|$ cluster around the same order of magnitude but exhibit visible scatter. 
The heatmap in Fig.~\ref{fig:epsL_tauplane} further highlights that the maximal CP violation occurs precisely within the neutrino-viable $\tau$ bands, confirming that, in this framework, the imaginary part of $\tau$ acts as the sole physical source of CP violation.

To illustrate the quantitative realization of perfect co-genesis, 
we select benchmark points (BP1–BP4) from the neutrino-oscillation consistent $\tau$ region,
each reproducing the observed baryon asymmetry and dark matter relic density within 
the $3\sigma$ NuFIT~5.2 limits. 
The corresponding parameters and cosmological observables are summarized in 
Table~\ref{tab:benchmarks}. 
These benchmarks serve as representative examples for the subsequent Boltzmann evolution.

\begin{figure}[htbp]
\centering
\begin{minipage}{0.48\textwidth}
\centering
\includegraphics[width=1.1\textwidth]{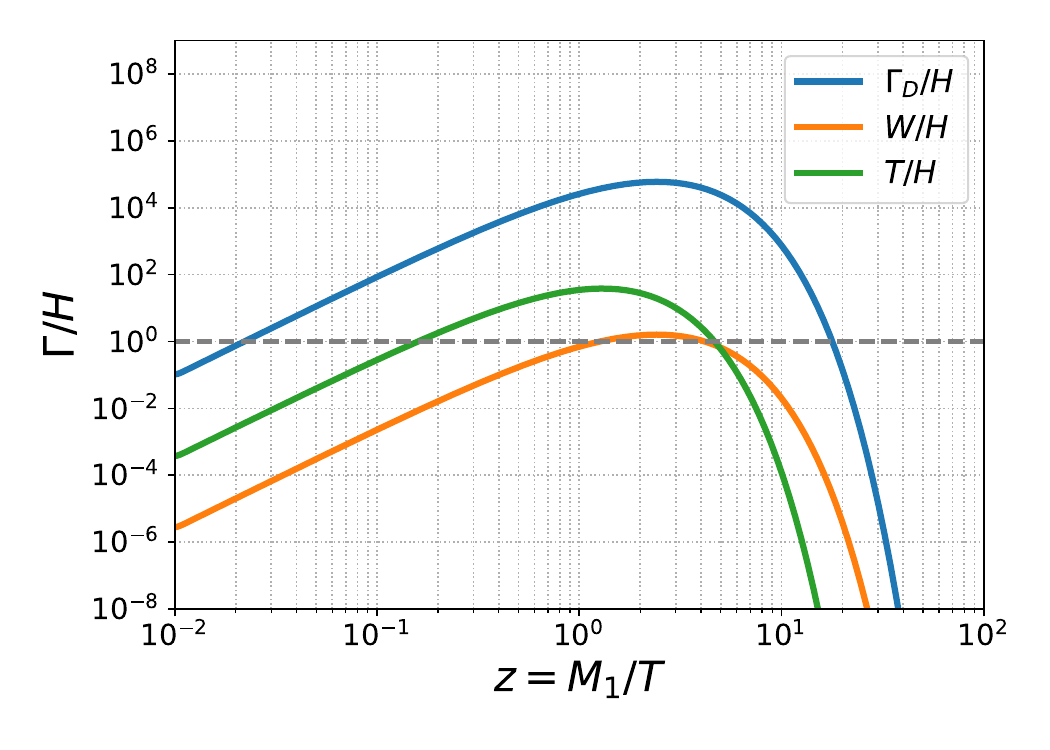}
\end{minipage}
\hfill
\begin{minipage}{0.48\textwidth}
\centering
\includegraphics[width=1.1\textwidth]{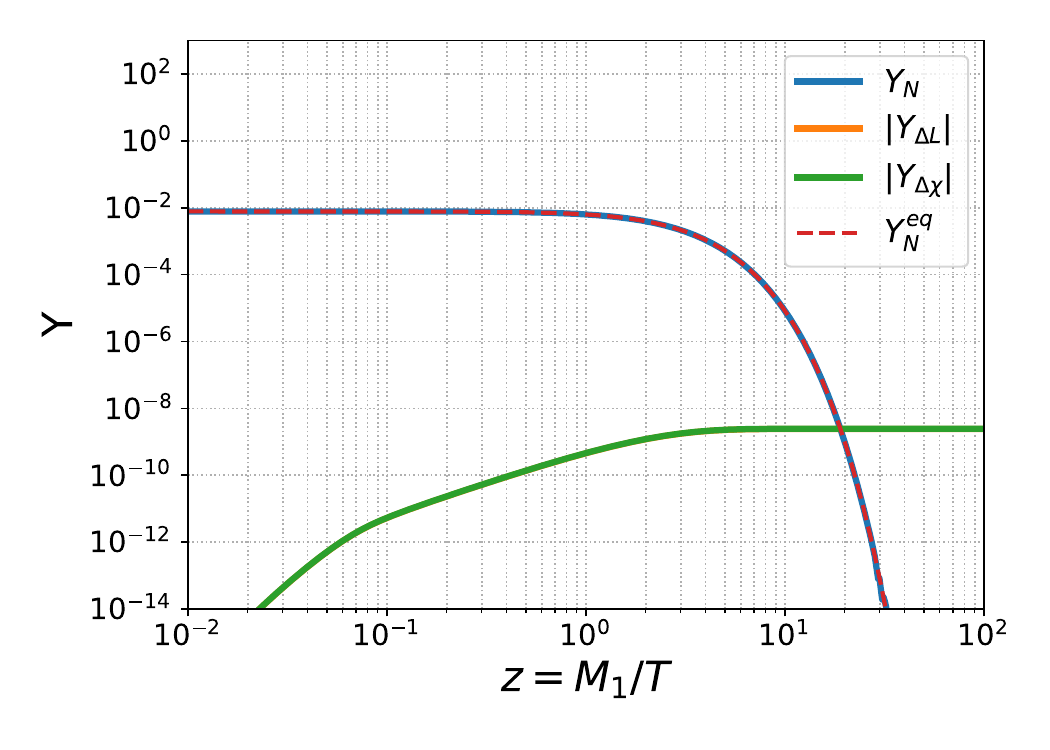}
\end{minipage}
\begin{minipage}{0.48\textwidth}
\centering
\includegraphics[width=1.1\textwidth]{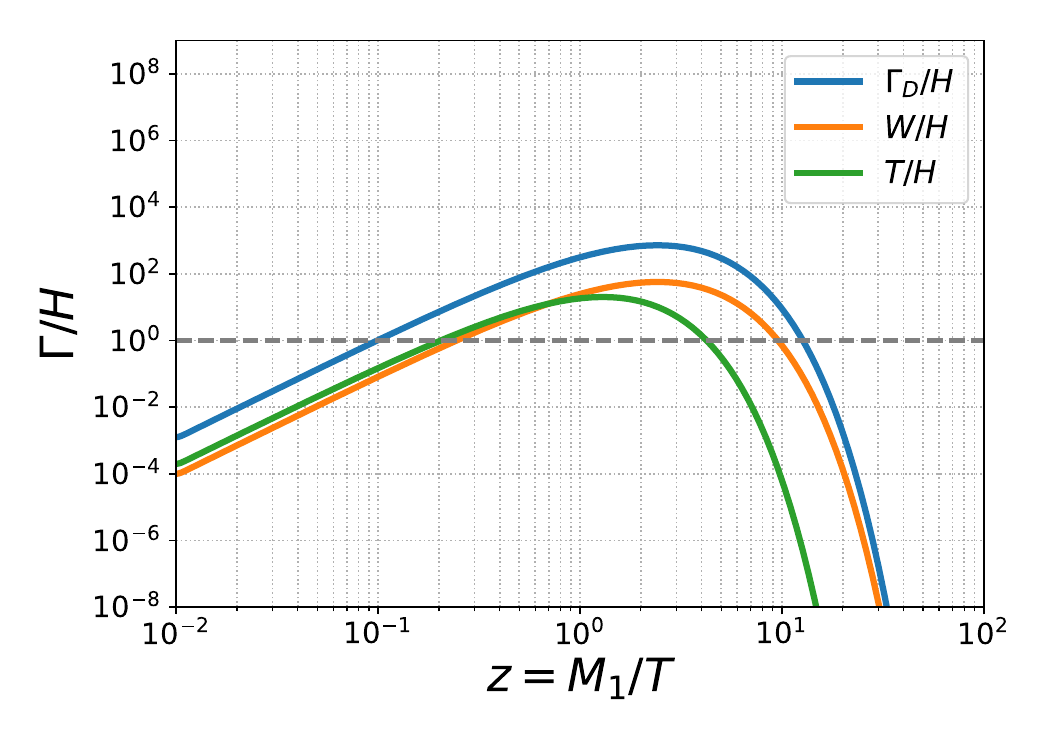}
\end{minipage}
\hfill
\begin{minipage}{0.48\textwidth}
\centering
\includegraphics[width=1.1\textwidth]{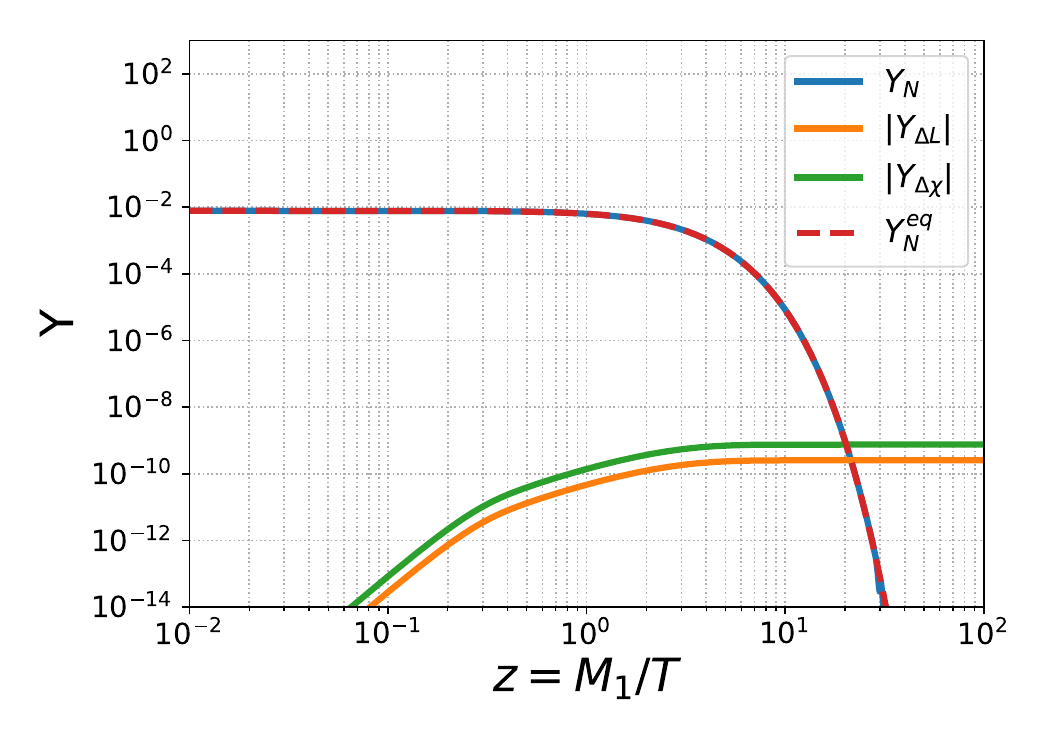}
\end{minipage}

\caption{
Boltzmann evolution for benchmark points BP4 (\textbf{upper row}, $y_{\mathrm{DM}}=10^{-2}$, balanced cogenesis) and BP1 (\textbf{lower row}, $y_{\mathrm{DM}}=10^{-3}$, dark-dominated). 
Left: thermally averaged decay, washout, and transfer rates as functions of $z=M_1/T$. 
Right: comoving yields of the heavy triplet and the lepton and dark asymmetries. 
Both asymmetries freeze out around $z\gtrsim6$, with co-genesis realized near intermediate $y_{\mathrm{DM}}$ values.
}
\label{fig:BoltzmannEvol}
\end{figure}
\begin{table}[htbp]
\centering
\caption{
Benchmark points illustrating the transition between dark- and balanced co-genesis regimes. All benchmarks reproduce the observed baryon asymmetry and dark relic density within uncertainties.}
\renewcommand{\arraystretch}{1.15}
\setlength{\tabcolsep}{5pt}
\footnotesize
\begin{tabular}{lccccccccc}
\hline
Point & $\tau$ & $M_1$ [$10^7$ GeV] & $|\epsilon_L|$ & $|\epsilon_\chi|$ &
$\eta_B$ & $\Omega_\chi h^2$ & $\Omega_\chi/\Omega_B$ & $m_\chi$ [GeV] \\
\hline
\multicolumn{9}{c}{$y_{\mathrm{DM}} = 10^{-3}$ \, (dark-dominated)} \\
\hline
BP1 & $1.37+i2.63$ & 1.9 & $2.3{\times}10^{-8}$ & $4.3{\times}10^{-7}$ & $4.6{\times}10^{-10}$ & 0.10 & 4.03 & 0.97 \\
BP2 & $1.64+i2.57$ & 2.1 & $3.05{\times}10^{-8}$ & $7.72{\times}10^{-8}$ & $5.9{\times}10^{-10}$ & 0.116 & 5.19 & 0.70 \\
\hline
\multicolumn{9}{c}{$y_{\mathrm{DM}} = 10^{-2}$ \, (balanced cogenesis)} \\
\hline
BP3 & $1.64+i2.52$ & 1.96 & $3.17{\times}10^{-8}$ & $3.19{\times}10^{-8}$ & $6.15{\times}10^{-10}$ & 0.121 & 5.4 & 1.78 \\
BP4 & $1.37+i2.69$ & 4.16 & $3.37{\times}10^{-8}$ & $3.39{\times}10^{-8}$ & $6.5{\times}10^{-10}$ & 0.12 & 5.7 & 1.76 \\
\hline
\end{tabular}
\label{tab:benchmarks}
\end{table}

The left panel presents the thermally averaged decay, washout, and transfer rates 
normalized to the Hubble expansion. 
At early times ($z\!=\!M_1/T<1$) the total decay rate $\Gamma_D/H$ 
is several orders of magnitude above unity, keeping $\Sigma_1$ in thermal equilibrium. 
As the Universe cools and $z$ approaches unity, the decay rate falls below the Hubble expansion, 
marking the onset of the out-of-equilibrium regime required by the Sakharov conditions. 
The washout ($W/H$) and transfer ($T/H$) processes remain much smaller than unity during this epoch, 
ensuring that the generated asymmetries are only mildly erased.

The Boltzmann evolution of the heavy triplet and the resulting asymmetries is shown in 
Fig.~\ref{fig:BoltzmannEvol} (right panel) for representative portal couplings 
$y_{\text{DM}} = 10^{-3}$ and $10^{-2}$, corresponding respectively to benchmark points 
BP1 (dark-dominated) and BP4 (balanced co-genesis) listed in Table~\ref{tab:benchmarks}. 
In both cases, the triplet remains in thermal equilibrium for $z < 1$ and departs near 
$z \simeq 1$, satisfying the out-of-equilibrium condition required for leptogenesis. 
Once the decays of $\Sigma_1$ become inefficient ($z \gtrsim 1$), CP-violating decays 
generate simultaneous lepton and dark asymmetries. The comoving yields $Y_N$, 
$|Y_{\Delta L}|$, and $|Y_{\Delta \chi}|$ evolve as shown in the right panels of 
Fig.~\ref{fig:BoltzmannEvol}: both asymmetries grow and freeze out around $z \gtrsim 6$, 
when $\Gamma_D/H \ll 1$. For small $y_{\text{DM}}$, the dark yield $|Y_{\Delta\chi}|$ 
dominates, while at larger $y_{\text{DM}}$ the visible and dark yields become comparable, 
reflecting their nearly equal CP asymmetries. The balanced co-genesis regime is therefore 
realized around $y_{\text{DM}} \sim 10^{-2}$, where $|Y_{\Delta L}^\infty| \simeq |Y_{\Delta\chi}^\infty|$. 
Other benchmark points exhibit qualitatively similar evolution with only mild variations 
in the final asymmetry ratios, so BP1 and BP4 sufficiently illustrate the generic behavior 
of the co-genesis dynamics in this framework.

\subsection{Correlation between CP Asymmetries and Relic Densities}

\begin{figure}[htbp]
\centering
\begin{minipage}{0.95\textwidth}
\centering
\includegraphics[width=0.48\textwidth]{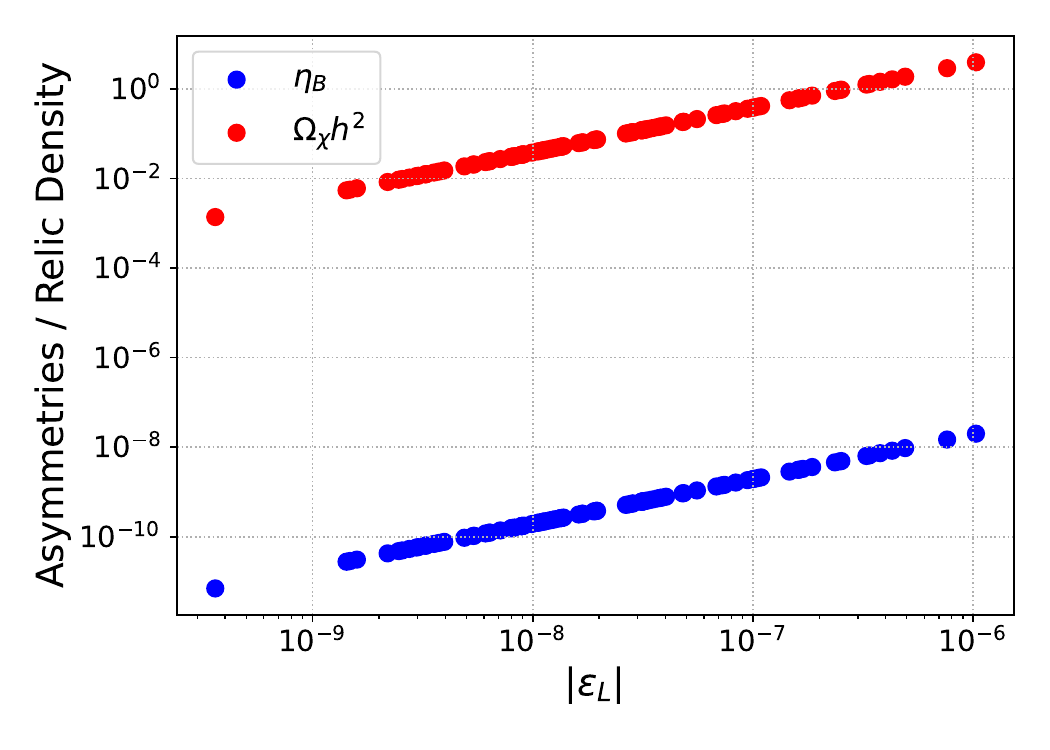}
\includegraphics[width=0.48\textwidth]{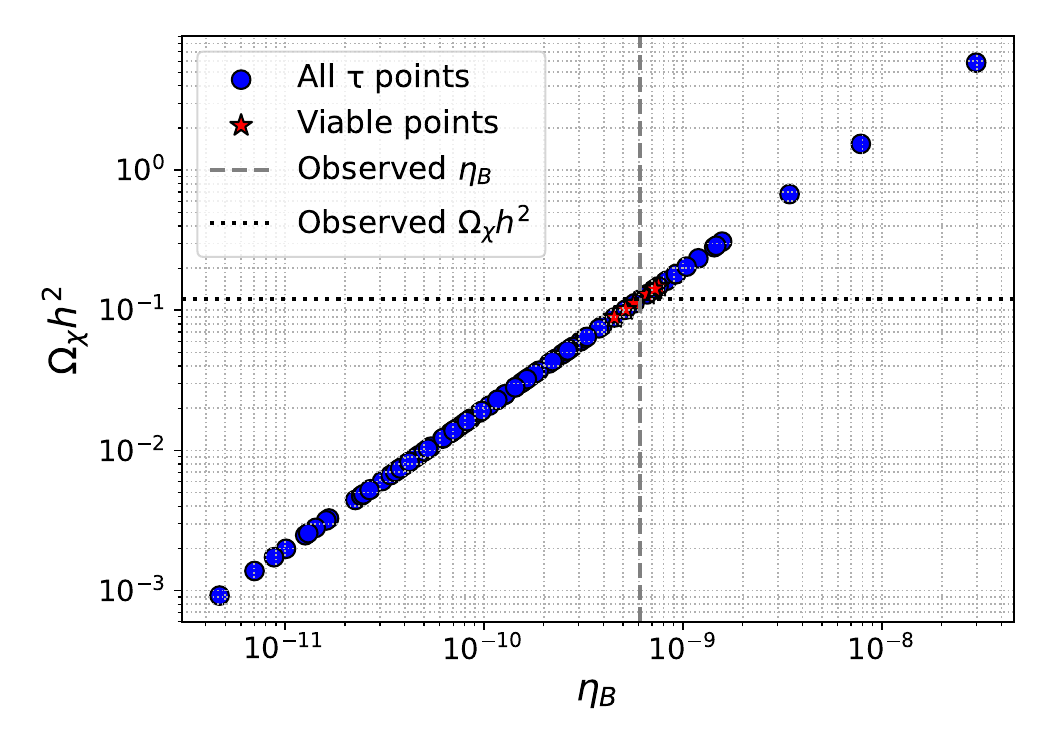}

\end{minipage}
\caption{
Correlation between microscopic CP asymmetries and macroscopic relic observables.
(a) Baryon asymmetry $\eta_B$ (blue) and dark matter relic density $\Omega_\chi h^2$ (red)
as functions of the CP asymmetry $|\epsilon_L|$,
showing linear scaling in the regime of unsaturated washout and mild saturation at larger values.
(b) Direct correlation between $\Omega_\chi h^2$ and $\eta_B$
for all $\tau$ points (blue) and neutrino-viable points (red stars),
demonstrating co-genesis near the observed values
$\eta_B^{\rm obs} = 6.12\times10^{-10}$ and $\Omega_\chi h^2 = 0.12$.
}
\label{fig:asymmetry_correlation}
\end{figure}

The solutions of the coupled Boltzmann equations~(\ref{eq:Boltz_Sigma})–(\ref{eq:Boltz_DM})
for the benchmark points in Table~\ref{tab:benchmarks}
yield the final asymmetries $Y_{\Delta L}^\infty$ and $Y_{\Delta\chi}^\infty$,
which are subsequently converted into the baryon and dark matter relic abundances
through Eqs.~(\ref{eq:sphaleron})–(\ref{eq:Omegachi})~\cite{Fong:2010qh}.
Fig.~\ref{fig:asymmetry_correlation} illustrates the numerical correlation
between microscopic CP asymmetries $(\epsilon_L,\epsilon_\chi)$
and macroscopic observables $\eta_B$ and $\Omega_\chi h^2$. Both balanced cogenesis and dark-sector dominated regimes are included in this combined $\tau$-scan.
Their contributions appear as a mild spread in the correlation plots, reflecting the small variations of the branching ratios and CP phases across the parameter space.

At fixed heavy-triplet mass $M_1$,
both $\eta_B$ and $\Omega_\chi h^2$ scale linearly with the corresponding CP asymmetries
for small $|\epsilon_{L,\chi}|$, as expected when washout is not yet saturated.
Beyond this range, partial washout effects flatten the growth of the generated asymmetries,
reflecting the onset of asymmetry damping.
Across the neutrino-viable $\tau$ region,
the baryon asymmetry consistently falls near the observed value
$\eta_B^{\rm obs} = (6.12 \pm 0.03)\times10^{-10}$~\cite{Planck:2018vyg},
while the dark relic density matches the Planck constraint
$\Omega_\chi h^2 = 0.12 \pm 0.001$,
demonstrating that both sectors reach the correct magnitudes
without additional parameter tuning.
Because $\epsilon_\chi$ tracks $\epsilon_L$ up to ${\cal O}(1)$ factors,
the linear trend with $|\epsilon_L|$ in Fig.~\ref{fig:asymmetry_correlation}(a)
effectively captures the behavior of both CP sources.

Numerically, the ratio $\Omega_\chi/\Omega_B$ stabilizes in a narrow band
$\Omega_\chi/\Omega_B \simeq 5$–$7$,
fully consistent with the cosmological value
$\Omega_\chi/\Omega_B \simeq 5.4$.
The mild spread arises from small ${\cal O}(1)$ differences
between the modular phases of $Y_\nu$ and $Y_\chi$,
confirming that the co-genesis correlation originates directly
from the modular flavor geometry rather than from any ad-hoc phase tuning.

Overall, the modular $S_4$ structure enforces a robust correlation at the order-of-magnitude level
between $\epsilon_L$, $\epsilon_\chi$, and the final relic abundances.
Since the same modulus $\tau$ governs both visible and dark CP violation,
the observed $\Omega_\chi/\Omega_B$ ratio emerges naturally from the modular geometry,
providing a unified and predictive framework for baryon–dark matter co-genesis.
Because $\epsilon_\chi$ tracks $\epsilon_L$ up to ${\cal O}(1)$ factors,
the linear trend with $|\epsilon_L|$ in Fig.~\ref{fig:asymmetry_correlation}
effectively captures the behavior of both CP sources. For clarity, we present the correlation only as a function of the visible-sector CP asymmetry $|\epsilon_L|$, since the dark-sector asymmetry $|\epsilon_\chi|$ exhibits the same behaviour due to its common modular origin and comparable magnitude throughout the viable parameter space.
\subsection{Dependence on Dark Matter Mass and Benchmark Predictions}

To further quantify the co-genesis mechanism, we analyze the dependence of
the dark matter relic density $\Omega_\chi h^2$
and the ratio $\Omega_\chi / \Omega_B$
on the dark fermion mass $m_\chi$.
The results are obtained by solving the Boltzmann equations
for the benchmark points in Table~\ref{tab:benchmarks},
keeping all other parameters fixed to their neutrino-viable modular values.

\begin{figure}[htbp]
\centering
\begin{minipage}{0.95\textwidth}
\centering
\includegraphics[width=0.48\textwidth]{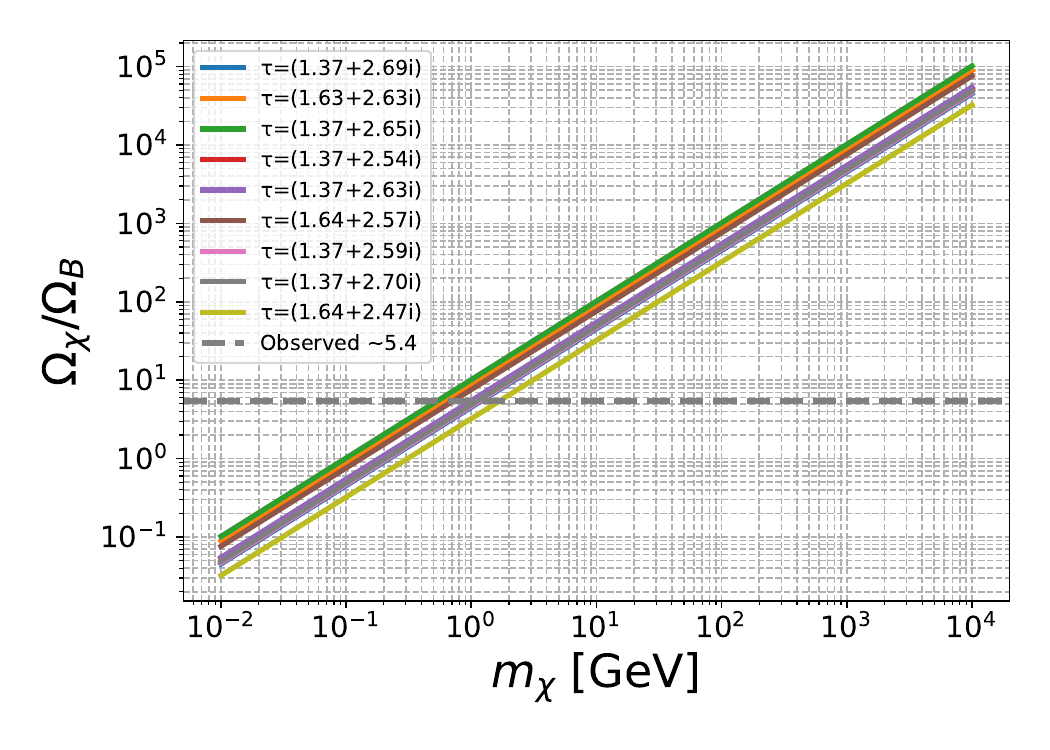}
\includegraphics[width=0.48\textwidth]{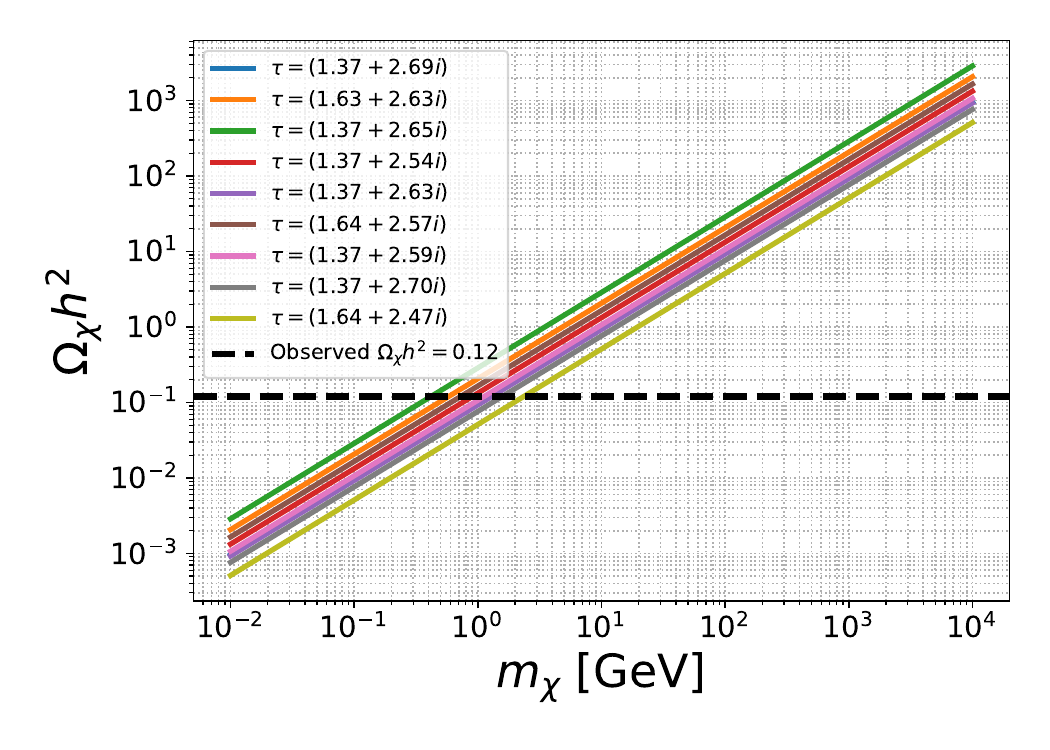}
\end{minipage}
\caption{
Dependence of the relic abundances on the dark matter mass $m_\chi$
for the benchmark points in Table~\ref{tab:benchmarks}.
Left: ratio $\Omega_\chi/\Omega_B$ as a function of $m_\chi$.
Right: dark matter relic density $\Omega_\chi h^2$ versus $m_\chi$.
Both show an approximately linear trend once the asymmetry is frozen,
and all benchmark trajectories nearly coincide,
reflecting the small spread of freeze-out asymmetries across the viable $\tau$ region.
}
\label{fig:Omega_mchi}
\end{figure}

Figure~\ref{fig:Omega_mchi} shows that both quantities scale approximately linearly with $m_\chi$ once the dark and visible asymmetries
$Y_{\Delta\chi}^\infty$ and $Y_{\Delta L}^\infty$ have frozen out.
In this regime, the relic abundance follows
$\Omega_\chi h^2 \propto m_\chi |Y_{\Delta\chi}^\infty|$,
and the ratio $\Omega_\chi/\Omega_B$ increases proportionally with $m_\chi$.
All benchmark trajectories nearly coincide,
indicating that the freeze-out asymmetries are very similar
throughout the neutrino-consistent $\tau$ region.
The intersection with the observed Planck values,
$\Omega_\chi h^2 = 0.120 \pm 0.001$ and
$\Omega_\chi/\Omega_B \simeq 5.4$~\cite{Planck:2018vyg},
occurs for
\begin{equation}
m_\chi \simeq (0.1\text{--}2)~\mathrm{GeV},
\end{equation}
in excellent agreement with the numerical benchmarks in
Table~\ref{tab:benchmarks}.

Because the same $\tau$-controlled CP asymmetries determine both
$Y_{\Delta L}^\infty$ and $Y_{\Delta\chi}^\infty$,
the preferred value $m_\chi \simeq 0.5$--$2~\mathrm{GeV}$
effectively encodes neutrino-sector information
in the dark matter mass prediction.
The resulting $m_\chi$ is therefore not imposed by hand,
but emerges naturally from the modular $S_4$ dynamics that govern
both visible and dark asymmetry generation.

\paragraph{Remarks on theoretical assumptions.}
The present analysis is based on the supersymmetric modular $S_4$ framework, so it is worth clarifying how the different energy scales remain consistent within the framework. 
The model is analyzed in the effective non-supersymmetric limit, yet it can naturally arise from a 
supersymmetric or supergravity completion in which the modulus $\tau$ and its associated modular 
functions $Y_i(\tau)$ originate from a chiral modulus field. Supersymmetry breaking may be represented 
by a spurion field $X = \theta^2 F$, which transmits soft terms of typical size 
$m_{\mathrm{SUSY}} \simeq F/M_{\mathrm{mess}}$, where $M_{\mathrm{mess}}$ is the mediation scale\cite{Feruglio:2017rjo}.

In the parameter range explored here, the heavy fermionic triplets have characteristic masses
\[
M_\Sigma \sim 10^{7}\,\mathrm{GeV}, 
\]
while the dark matter mass remains in the few--GeV regime. A plausible and phenomenologically 
stable hierarchy of
\[
m_{\mathrm{SUSY}} \sim 10^{6\text{--}8}\,\mathrm{GeV}, \qquad 
M_{\mathrm{mess}} \sim 10^{14\text{--}16}\,\mathrm{GeV},
\]
leads to a ratio $m_{\mathrm{SUSY}}/M_{\mathrm{mess}} \lesssim 10^{-8}$, which keeps any 
supersymmetric threshold effects on the modular Yukawa structure extremely small.

Because of this suppression, the modular flavor relations remain radiatively stable, and the 
predictions derived from them are not distorted by higher-scale corrections. Furthermore, all 
possible superpartners are expected to decouple well before the leptogenesis epoch, leaving a 
consistent low-energy theory that contains only the Standard Model fields, the fermionic 
triplets $\Sigma_i$, and the dark-sector fields $(\chi, \phi)$. This ensures that the effective 
description used in our analysis remains valid throughout the relevant cosmological evolution.

\paragraph{Time dependence of the CP asymmetry.}
In resonant leptogenesis, a full quantum treatment reveals that the CP asymmetry 
$\epsilon(t)$ can exhibit oscillatory time dependence when the mass splitting 
$\Delta M$ becomes comparable to the decay width $\Gamma$~\cite{Flanz:1996nh}. 
In the present analysis, we employ the standard momentum-averaged Boltzmann 
formalism~\cite{Buchmuller:2002rq}, using time-independent asymmetries 
$\epsilon_{L,\chi}$ derived from the one-loop decay amplitudes. 
For the benchmark parameters $\Delta M/M_1 = 10^{-3}–10^{-5}$ and 
$\Gamma/M_1 \sim 10^{-6}$, the coherence time of the quasi-degenerate states 
is much shorter than the timescale of asymmetry generation, 
so any oscillations in $\epsilon(t)$ are effectively averaged out. 
Consequently, the time-averaged asymmetry provides an accurate estimate of the 
integrated baryon yield~\cite{Garbrecht:2013psa}. Comparisons with quantum-kinetic treatments indicate that deviations from the
time-independent approximation remain at the level of at most
$\mathcal{O}(10\%)$ in this regime. More explicitly, the literature on resonant 
leptogenesis identifies three distinct physical 
regimes depending on the ratio $\Delta M/\Gamma$:
the weakly resonant regime ($\Delta M \gg \Gamma$) 
where the static Boltzmann treatment is accurate 
and insensitive to the regulator form, the 
strongly resonant regime ($\Delta M \sim \Gamma$) 
where quantum coherence effects are important, 
and the maximally resonant regime 
($\Delta M \lesssim \Gamma$) where the Boltzmann 
treatment is qualitatively correct at best. 
This hierarchy has been explicitly established 
by full CTP quantum kinetic 
treatments~\cite{Jukkala:2021}.

In our parameter space, $\Delta M/M_1 = 
10^{-3}$--$10^{-5}$ and $\Gamma/M_1 \sim 
10^{-6}$, giving:
\begin{equation}
\frac{\Delta M}{\Gamma} = 
\frac{\Delta M/M_1}{\Gamma/M_1} = 
\frac{10^{-3}\text{--}10^{-5}}{10^{-6}} = 
10^3\text{--}10 \gg 1.
\end{equation}
Our entire parameter space therefore lies 
firmly in the weakly resonant regime, where 
the static Boltzmann treatment with the 
Pilaftsis--Underwood regulator is explicitly 
confirmed to be accurate and insensitive to 
the precise form of the regulator~\cite{Jukkala:2021}. 
The strongly resonant regime $\Delta M \sim \Gamma$ 
is never reached in our parameter space.

\paragraph{Flavour effects.}
Flavour-dependent washout effects can in principle become relevant for $M_{1,2} \lesssim 10^{12}~\mathrm{GeV}$~\cite{Abada:2006ea,Nardi:2006fx}. At the scale $M_1 \sim 10^7~\mathrm{GeV}$ relevant for our analysis, charged-lepton Yukawa interactions are in equilibrium and flavour effects may therefore play a role. However, the situation in the type-III seesaw is qualitatively different from the type-I case, as established in the dedicated study of Aristizabal Sierra, Kamenik and Nemev\v{s}ek~\cite{AristizabalSierra:2010mv}. The fermionic $\mathrm{SU}(2)_L$ triplets $\Sigma_i$ carry electroweak gauge quantum numbers and interact with all lepton doublets through both Yukawa interactions and flavour-blind $\mathrm{SU}(2)_L$ gauge scatterings. In this framework, gauge reactions tend to suppress flavour-dependent differences in the washout dynamics, so that flavour effects can be reduced for the lightest triplet responsible for generating the asymmetry~\cite{AristizabalSierra:2010mv}.

In our specific modular $S_4$ framework, this behaviour is further supported by the structure of the Dirac Yukawa matrix $Y_\nu$, constructed from the weight-4 modular forms $Y_3^{(4)}$ and $Y_{3'}^{(4)}$ [Eqs.~(2.5)--(2.6)]. For the phenomenologically viable values of the modulus parameter $\tau$ ($\mathrm{Re}\,\tau \simeq 1.37$ and $1.64$, $\mathrm{Im}\,\tau \simeq 2.6$--$2.8$), these modular forms yield flavour projectors
\[
P_{i\alpha} =
\frac{\Gamma(\Sigma_i \to \ell_\alpha H)}
{\Gamma(\Sigma_i \to L H)}
\]
satisfying
\[
|P_{i,e}| \sim |P_{i,\mu}| \sim |P_{i,\tau}|
\]
within $\mathcal{O}(1)$ factors, so that no single flavour channel dominates the washout dynamics. For such approximately democratic flavour projectors, the single-flavour Boltzmann treatment is expected to capture the total $B-L$ asymmetry up to $\mathcal{O}(1)$ corrections~\cite{AristizabalSierra:2010mv}. A fully flavour-resolved treatment would mainly redistribute the asymmetry among flavours without qualitatively affecting the co-genesis conclusions.

\section{Conclusion}
\label{sec:res}
In this work we have presented a unified framework for neutrino masses, baryogenesis,
and dark matter based on a modular $S_4$ symmetry with a type-III seesaw mechanism.
The model simultaneously explains the observed neutrino oscillation data and
the cosmological baryon and dark matter relic abundances through a common origin in the
complex modulus~$\tau$, which determines all Yukawa couplings and CP phases.

A detailed numerical analysis was performed for the normal hierarchy of neutrino masses.
The modular structure naturally restricts $\tau$ to narrow viable bands around
$\mathrm{Re}\,\tau\simeq1.35$ and $1.64$ with $\mathrm{Im}\,\tau\simeq2.6$–$2.8$,
where the modular forms reproduce the observed mixing angles and mass-squared
differences within the $3\sigma$ ranges of the NuFIT~5.2 (2024) global fit.
The framework predicts a nearly maximal leptonic CP phase
$\delta_{\rm CP}\!\simeq\!\pm(150^\circ\!-\!180^\circ)$
and an effective Majorana mass
$m_{\beta\beta}\!\simeq\!(8$–$18)\!\times\!10^{-3}$~eV,
which lies below the present KamLAND-Zen bound but within the projected sensitivity
of nEXO and LEGEND-1000.

The same modular parameter~$\tau$ that controls the neutrino sector also fixes the
Yukawa couplings governing the CP-violating decays of the heavy triplet fermions,
thereby linking the visible and dark sectors.
The resulting CP asymmetries $|\epsilon_{L,\chi}|\!\sim\!10^{-9}$–$10^{-6}$
generate comparable lepton and dark asymmetries through resonant decays of
$\Sigma_{1,2}$ at $M_{1,2}\!\sim\!10^{7}$~GeV.
Solutions of the coupled Boltzmann equations show that both asymmetries freeze out
with similar magnitudes, yielding the observed baryon asymmetry
$\eta_B\simeq6\times10^{-10}$ and the correct dark relic density
$\Omega_\chi h^2\simeq0.12$ without any parameter tuning.

The co-genesis mechanism exhibits a strong numerical correlation between
$\eta_B$ and $\Omega_\chi h^2$, reflecting the common modular origin of
the CP asymmetries.
The ratio $\Omega_{\chi} / \Omega_{\mathrm{B}}$ stabilizes near the cosmological value
$3 \lesssim \Omega_{\chi} / \Omega_{\mathrm{B}} \lesssim 7$, corresponding to a dark-matter
mass prediction $m_{\chi} \simeq 0.1\text{--}2\ \text{GeV}$.
The predicted GeV-scale dark-matter mass and its extremely suppressed interactions
with the visible sector ensure full consistency with existing laboratory,
astrophysical, and cosmological constraints on asymmetric fermionic dark matter.

The predictive power of the modular $S_4$ symmetry lies in its ability to relate
flavor, CP violation, and cosmology through the single modulus~$\tau$.
Once $\tau$ is fixed by low-energy neutrino data,
all other observables—baryon asymmetry, dark matter abundance, and
$0\nu\beta\beta$ predictions—follow without additional parameters.
This work therefore demonstrates a fully predictive realization of
baryon–dark matter co-genesis embedded in a modular flavor theory,
linking the origin of flavor and matter–antimatter asymmetry through a
common geometric parameter.
Future measurements of the leptonic CP phase,
neutrinoless double beta decay, and searches for light asymmetric dark matter
will provide decisive tests of this unified modular framework.

\appendix
\section*{Appendix~A: Essentials of Modular Symmetry and $S_4$ Yukawa Multiplets}
\label{app:A}
\addcontentsline{toc}{section}{Appendix~A: Essentials of Modular Symmetry and $S_4$ Yukawa Multiplets}
\renewcommand{\theequation}{A.\arabic{equation}}

Modular symmetry provides a geometric approach to flavor symmetries in which
Yukawa couplings are expressed as modular forms—holomorphic functions of a single
complex parameter~$\tau$ in the upper half-plane ($\mathrm{Im}\,\tau>0$).
The modulus transforms under the modular group $\mathrm{SL}(2,\mathbb{Z})$ as
\begin{equation}
\tau \longrightarrow \frac{a\tau+b}{c\tau+d},
\qquad
a,b,c,d\in\mathbb{Z},\quad ad-bc=1.
\end{equation}
The two generators,
\begin{equation}
S:\ \tau\mapsto -\frac{1}{\tau},
\qquad
T:\ \tau\mapsto \tau+1,
\end{equation}
satisfy $S^2=(ST)^3=\mathbb{I}$ and define the modular group relations.
Finite subgroups $\Gamma_N=\overline{\Gamma}(1)/\overline{\Gamma}(N)$ correspond
to well-known discrete flavor symmetries:
\[
\Gamma_2\simeq S_3,\qquad
\Gamma_3\simeq A_4,\qquad
\Gamma_4\simeq S_4,\qquad
\Gamma_5\simeq A_5.
\]
In this work we employ $\Gamma_4\simeq S_4$, 
the smallest modular group that yields realistic lepton mixing while
retaining a nontrivial structure of modular forms.

A modular form of weight~$k$ and level~$N$ transforms as
\begin{equation}
Y_i(\tau)\;\longrightarrow\;
(c\tau+d)^k\,\rho_{ij}(\gamma)\,Y_j(\tau),
\qquad
\gamma=\begin{pmatrix}a&b\\c&d\end{pmatrix}\!\in\!\Gamma_N,
\end{equation}
where $\rho(\gamma)$ is a unitary representation of the finite modular group.
The number of independent modular forms of weight~$2k$ increases with~$N$,
as summarized in Table~\ref{tab:modular_form_count}.
\begin{table}[htbp]
\centering
\caption{Number of linearly independent modular forms of weight $2k$ for low modular levels~$N$.}
\begin{tabular}{|c|c|c|}
\hline
Level $N$ & Associated group & $\#$ of forms (weight $2k$) \\ \hline
2 & $S_3$ & $k+1$ \\
3 & $A_4$ & $2k+1$ \\
4 & $S_4$ & $4k+1$ \\
5 & $A_5$ & $10k+1$ \\ \hline
\end{tabular}
\label{tab:modular_form_count}
\end{table}

---

\subsection*{A.1 Structure of the $S_4$ Group}

The finite group $S_4$ contains five irreducible representations:
$\mathbf{1}$, $\mathbf{1'}$, $\mathbf{2}$, $\mathbf{3}$, and $\mathbf{3'}$.
Its generators satisfy
\begin{equation}
S^2=T^4=(ST)^3=\mathbb{I}.
\end{equation}
The relevant Clebsch–Gordan (CG) products are
\begin{align}
1'\!\otimes\!1' &= 1, &
2\!\otimes\!2 &= 1\oplus1'\oplus2, &
3\!\otimes\!3 &= 1\oplus2\oplus3\oplus3', \nonumber\\
2\!\otimes\!3 &= 3\oplus3', &
3\!\otimes\!3' &= 1'\oplus2\oplus3\oplus3'. &&
\end{align}
Explicit CG coefficients and representation matrices are listed in
Refs.~\cite{Novichkov:2020uzt,Wang:2020}
and are used to build modular-invariant Yukawa interactions.

\subsection*{A.2 Modular Forms under $S_4$}

At level~$N=4$, the lowest-weight modular forms furnish a doublet
$Y^{(2)}_{\mathbf{2}}=(Y_1,Y_2)^{T}$ and a triplet-prime
$Y^{(2)}_{\mathbf{3'}}=(Y_3,Y_4,Y_5)^{T}$.
With $q\equiv e^{2\pi i\tau}$, their schematic $q$-expansions are
\begin{align}
Y_1 &\simeq 1 + 3q + 3q^2 + \mathcal{O}(q^3), &
Y_2 &\simeq \sqrt{3}\,q^{1/2}(1 + 4q + 6q^2 + \cdots), \nonumber\\
Y_3 &\simeq 1 - 2q + 6q^2 + \cdots, &
Y_4 &\simeq \sqrt{2}\,q^{1/4}(1 + 6q + 13q^2 + \cdots), \nonumber\\
Y_5 &\simeq 4\sqrt{2}\,q^{3/4}(1 + 2q + 3q^2 + \cdots). &&
\end{align}
Higher-weight modular multiplets are constructed by modular-covariant
products of these basic forms.  
The independent weight-four combinations used in our model are
\begin{align}
Y^{(4)}_{\mathbf{1}} &= Y_1^2 + Y_2^2, &
Y^{(4)}_{\mathbf{2}} &= (Y_2^2 - Y_1^2,\; 2Y_1Y_2)^{T}, \nonumber\\[2pt]
Y^{(4)}_{\mathbf{3}} &= (-2Y_2Y_3,\;
 \sqrt{3}Y_1Y_5 + Y_2Y_4,\;
 \sqrt{3}Y_1Y_4 + Y_2Y_5)^{T}, \nonumber\\[2pt]
Y^{(4)}_{\mathbf{3'}} &= (2Y_1Y_3,\;
 \sqrt{3}Y_2Y_5 - Y_1Y_4,\;
 \sqrt{3}Y_2Y_4 - Y_1Y_5)^{T}.
\end{align}
The superscript ``(4)'' denotes the total modular weight.
These multiplets generate all Yukawa couplings appearing in the
charged-lepton, neutrino, and dark sectors, ensuring modular-invariant
interactions consistent with $S_4$ representation theory.
In practice, only the relative magnitudes and phases of the
modular forms—fixed by the chosen value of $\tau$—enter physical observables. The group-theoretical structures, generator matrices, and modular-form expansions presented here follow the conventions established in \cite{Zhang:2021}.

\section*{Appendix~B: Standard Parameterization of Neutrino Observables}
\label{app:B}
\addcontentsline{toc}{section}{Appendix~B: Standard Parameterization of Neutrino Observables}
\renewcommand{\theequation}{B.\arabic{equation}}

The lepton mixing matrix $U_{\mathrm{PMNS}}$ is obtained via
\begin{equation}
U_{\mathrm{PMNS}} = U_\ell^\dagger U_\nu,
\end{equation}
where $U_\ell$ and $U_\nu$ diagonalize the charged-lepton and
neutrino mass matrices:
\begin{align}
U_\ell^\dagger M_\ell M_\ell^\dagger U_\ell &= 
{\rm diag}(m_e^2,m_\mu^2,m_\tau^2), &
U_\nu^\dagger M_\nu U_\nu^* &= 
{\rm diag}(m_1,m_2,m_3).
\end{align}

In the PDG convention,
\begin{equation}
U_{\mathrm{PMNS}} =
\begin{pmatrix}
1 & 0 & 0\\
0 & c_{23} & s_{23}\\
0 & -s_{23} & c_{23}
\end{pmatrix}
\!\begin{pmatrix}
c_{13} & 0 & s_{13}e^{-i\delta}\\
0 & 1 & 0\\
-s_{13}e^{i\delta} & 0 & c_{13}
\end{pmatrix}
\!\begin{pmatrix}
c_{12} & s_{12} & 0\\
-s_{12} & c_{12} & 0\\
0 & 0 & 1
\end{pmatrix}
P_\nu,
\end{equation}
where $c_{ij}\equiv\cos\theta_{ij}$, $s_{ij}\equiv\sin\theta_{ij}$,
and $P_\nu={\rm diag}(1,e^{i\alpha_{21}/2},e^{i\alpha_{31}/2})$.

Mixing observables follow as
\begin{align}
\sin^2\theta_{13} &= |U_{e3}|^2, &
\sin^2\theta_{12} &= \frac{|U_{e2}|^2}{1-|U_{e3}|^2}, &
\sin^2\theta_{23} &= \frac{|U_{\mu3}|^2}{1-|U_{e3}|^2},
\end{align}
and the CP-violating quantities are
\begin{align}
\delta_{\rm CP} &=
 -\arg\!\left[U_{e1}U_{e3}^*U_{\mu3}U_{\mu1}^*\right], &
J_{\rm CP} &= 
 \Im\!\left[U_{e1}U_{\mu2}U_{e2}^*U_{\mu1}^*\right].
\end{align}
The Majorana phases can be expressed as
\begin{align}
\alpha_{21} &= 2\,\arg\!\left(\frac{U_{e2}}{U_{e1}}\right), &
\alpha_{31} &= 2\,\arg\!\left(\frac{U_{e3}}{U_{e1}}\right)+2\delta_{\rm CP}.
\end{align}
The mass-squared splittings and total neutrino mass are
\begin{align}
\Delta m_{21}^2 &= m_2^2-m_1^2, &
\Delta m_{31}^2 &= m_3^2-m_1^2, &
\Sigma m_\nu &= m_1+m_2+m_3.
\end{align}

\section*{Appendix~C: Boltzmann Framework and Scattering Integrals}
\label{app:C}
\addcontentsline{toc}{section}{Appendix~C: Boltzmann Framework and Scattering Integrals}
\renewcommand{\theequation}{C.\arabic{equation}}

For completeness, we summarize the Boltzmann equation used in the two–sector analysis.
The comoving asymmetries 
$Y_{\Delta a}=Y_a-Y_{\bar a}$ for $a=L,\chi$ evolve according to~\cite{Falkowski:2011zr}
\begin{align}
\frac{dY_{\Delta a}}{dz}
&= - \frac{\Gamma_{\Sigma_1}}{H_1}\,
    \frac{z K_1(z)}{K_2(z)}
    \left( Y_{\Sigma_1}-Y_{\Sigma_1}^{\mathrm{eq}} \right)
    \epsilon_a
   - 2\,{\rm Br}_a^2\,I_W(z)\,Y_{\Delta a} \nonumber\\
&\quad
   - {\rm Br}_a {\rm Br}_b
     \big[ I_T^+(z)\,(Y_{\Delta a}+Y_{\Delta b})
         + I_T^-(z)\,(Y_{\Delta a}-Y_{\Delta b}) \big],
\label{eq:C1}
\end{align}
where $b\neq a$, $z=M_{_1}/T$, and $H_1=H(T=M_{_1})$.
The first term describes the CP–violating source from $\Sigma_1$ decays,
while the remaining terms account for washout and inter–sector transfer effects.

The thermal integrals entering Eq.~(\ref{eq:C1}) are defined as
\begin{align}
I_i(z) &= 
\frac{\hat{\Gamma}}{\pi} 
\int_0^{\infty} dt\, t^2\, K_1(t)\, f_i(t^2/z^2),
\end{align}
where $\hat{\Gamma} = \Gamma_{N_1}/M_{N_1}$, and
\begin{align}
f_W(s) &= 
\frac{s/2}{(s - 1)^2 + \hat{\Gamma}^2}
+ \frac{s - \log(s + 1)}{s}
- \frac{2(s - 1)}{(s - 1)^2 + \hat{\Gamma}^2}
  \frac{(s + 1)\log(s + 1) - s}{s} \notag \\
&\quad
+ \frac{s/2}{s + 1}
+ \frac{\log(s + 1)}{s + 2},
\\[6pt]
f_{T_+}(s) &= 
\frac{s/2}{(s - 1)^2 + \hat{\Gamma}^2}
+ \frac{s}{s + 1}
+ \frac{s - \log(s + 1)}{s},
\\[6pt]
f_{T_-}(s) &= 
\frac{s^2/2}{(s - 1)^2 + \hat{\Gamma}^2}
+ \frac{(s + 1)\log(s + 1) - s}{s + 1}
+ \frac{(s + 2)\log(s + 1) - 2s}{s}.
\end{align}

Here $I_W$ corresponds to $2\!\leftrightarrow\!2$ washout within a sector,
and $I_T^{\pm}$ encode transfer scatterings between sectors.

As $z = M_1/T$ increases, the pole (on-shell) contributions to the integrals are
exponentially suppressed, and inverse decays become subdominant.
In this regime, the Boltzmann evolution is controlled by off-shell
$2\leftrightarrow2$ scattering processes, which admit simple analytic
approximations.
For $z \gg 1$, the integrals reduce to the asymptotic forms
\begin{equation}
I_D(z) \simeq \frac{3W}{z^2}\,\frac{H_1}{\Gamma_{N_1}}, \qquad
I_{T^+}(z) \simeq \frac{W}{z^2}\,\frac{H_1}{\Gamma_{N_1}}, \qquad
I_{T^-}(z) \simeq \frac{14W}{z^4}\,\frac{H_1}{\Gamma_{N_1}},
\label{eq:asymptotic_integrals}
\end{equation}
where $W$ denotes the reduced scattering rate.

\section*{Appendix~D: Behavior of the CP Asymmetry Ratio as a Function of $y_{\mathrm{DM}}$}
\label{app:yDM_dependence}
\addcontentsline{toc}{section}{Appendix~D: Dependence on the Dark Yukawa Coupling $y_{\mathrm{DM}}$}
\begin{figure}[htbp]
\centering
\includegraphics[width=0.5\textwidth]{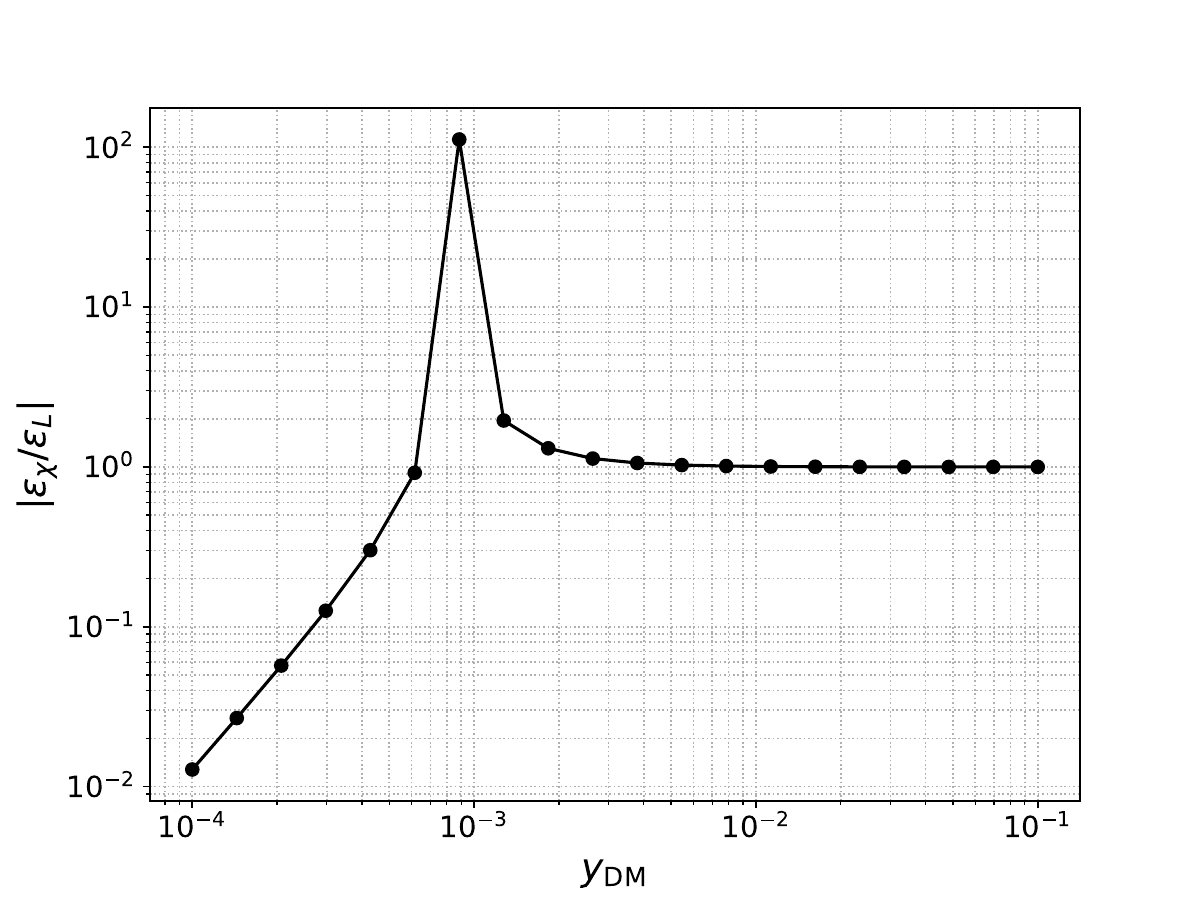}
\caption{Ratio of CP asymmetries $|\epsilon_\chi/\epsilon_L|$ as a function of the dark Yukawa coupling $y_{\mathrm{DM}}$ for a fixed representative value of the modulus $\tau = 1.64 + i\,2.57$. The horizontal dashed line indicates equal visible and dark CP asymmetries.}
\label{fig:YDM}
\end{figure} 
In order to clarify the origin of the dark-sector dominance observed at small values of $y_{\mathrm{DM}}$, we present in Fig.~\ref{fig:YDM} the ratio $|\epsilon_\chi/\epsilon_L|$ as a function of $y_{\mathrm{DM}}$ for a representative benchmark point in the modulus,
\[
\tau = 1.64 + i\,2.57.
\]

For $y_{\mathrm{DM}} \lesssim 10^{-3}$, the ratio exhibits a mild enhancement, arising from the different interference structures entering the visible and dark CP asymmetries. As $y_{\mathrm{DM}}$ increases, the ratio smoothly approaches unity, indicating comparable visible and dark contributions. This behavior confirms that the enhancement at small $y_{\mathrm{DM}}$ is a controlled quantitative effect rather than a breakdown of the perturbative expansion.

\bibliographystyle{unsrt}

\end{document}